\documentclass{article}
\usepackage{PRIMEarxiv}
\usepackage[utf8]{inputenc} 
\usepackage[T1]{fontenc}    
\usepackage{lmodern}
\usepackage{newtxtext}
\usepackage{hyperref}       
\usepackage{url}            
\usepackage{booktabs}       
\usepackage{amsfonts}       
\usepackage{nicefrac}       
\usepackage{microtype}      
\usepackage{lipsum}
\usepackage{fancyhdr}       
\usepackage{graphicx}       
\graphicspath{{media/}}     
\usepackage{multirow}%
\usepackage{amsmath,amssymb,amsfonts}%
\usepackage{amsthm}%
\usepackage{mathrsfs}%
\usepackage[title]{appendix}%
\usepackage{xcolor}%
\usepackage{textcomp}%
\usepackage{manyfoot}%
\usepackage{booktabs}%
\usepackage{algorithm}%
\usepackage{algorithmicx}%
\usepackage{algpseudocode}%
\usepackage{listings}%
\usepackage[utf8]{inputenc} 
\usepackage[T1]{fontenc}    

\usepackage{url}            
\usepackage{booktabs}       
\usepackage{amsfonts}       
\usepackage{nicefrac}       
\usepackage{microtype}      
\usepackage{lipsum}
\usepackage{graphics}
\usepackage[export]{adjustbox}
\usepackage{braket}
\usepackage{mathrsfs}
\usepackage{amsmath}
\usepackage{lmodern}
\usepackage{import}
\usepackage{subcaption}
\usepackage{caption}

\usepackage{bm}
\usepackage{xcolor} 
\usepackage{tikz} 
\usepackage{pgf}
\usepackage{pgfplots}
\usepgfplotslibrary{external}
\pgfplotsset{width=10cm,compat=1.9}
\title{Brain-Inspired Quantum Neural Architectures for Pattern Recognition: Integrating QSNN and QLSTM 
}

\author{
  Eva Andrés, Manuel Pegalajar Cuéllar, Gabriel Navarro \\
  Department of Computer Science and Artificial Intelligence \\
  University of Granada \\
  C/ Periodista Daniel Saucedo Aranda, s/n, 18071, Granada, Spain \\
  \texttt{\ e.evaandres@go.ugr.es, manupc@decsai.ugr.es, navarro@decsai.ugr.es} \\
}

\begin{document}
\maketitle

\begin{abstract}
Recent advances in the fields of deep learning and quantum computing have paved the way for innovative developments in artificial intelligence. In this manuscript, we leverage these cutting-edge technologies to introduce a novel model that emulates the intricate functioning of the human brain, designed specifically for the detection of anomalies such as fraud in credit card transactions. Leveraging the synergies of Quantum Spiking Neural Networks (QSNN) and Quantum Long Short-Term Memory (QLSTM) architectures, our approach is developed in two distinct stages, closely mirroring the information processing mechanisms found in the brain's sensory and memory systems. In the initial stage, similar to the brain's hypothalamus, we extract low-level information from the data, emulating sensory data processing patterns. In the subsequent stage, resembling the hippocampus, we process this information at a higher level, capturing and memorizing correlated patterns. We will compare this model with other quantum models such as Quantum Neural Networks among others and their corresponding classical models.
\end{abstract}


\section{Introduction}
In recent years, the field of artificial intelligence has witnessed remarkable advancements. A notable outcome of this progress is the emergence of brain-inspired artificial intelligence, a field that amalgamates insights from neuroscience, psychology, and computer science to develop more efficient and powerful AI systems \cite{WhenbraininspiredAImeetsAGI}. Additionally, quantum machine learning has emerged as a compelling area of interest within the scientific community. Notably, Variational Quantum Circuits (VQC) also know as Quantum Neural Networks (QNN) have demonstrated significant success across diverse domains, including supervised learning \cite{VQA-Classifiers, VQA-Neighbor, VQA-SVM} and unsupervised learning \cite{VQA_Unsupervised}. While research on Quantum Reinforcement Learning (QRL) is still in its early stages, recent studies have demonstrated that QNNs can outperform classical models in Reinforcement learning (RL) scenarios, achieving better cumulative rewards with fewer parameters to be learned \cite{QRL-energies}.

In classical machine learning, challenges persist when working with imbalanced datasets, limited data availability, or low data quality. Our aim is to investigate whether quantum models can yield improved results under such conditions. Moreover, we seek to explore the intriguing question of whether Quantum Brain-Inspired Models are capable of detecting patterns that elude classical and QNN models, potentially leading to enhanced learning outcomes and improved false positive and false negative rates.

Drawing inspiration from a theory of prefrontal cortex and hippocampus function, the hippocampus plays a role in the formation and retrieval of specific memories. Meanwhile, the prefrontal cortex accumulates features of related memories, shaping the 'context' of interconnected experiences. This context might encompass details like a list in which a set of words appeared, a common location for multiple events, or a shared set of ongoing task rules governing memory-related decisions. Consequently, when cued to a particular context, the prefrontal cortex influences the retrieval of memories pertinent to a context within the hippocampus and other brain regions \cite{prefrontalcortex}.

And finally, motivating this study we consider the wildly accepted memory model of Atkinson and Shifrin \cite{memory}, including short-term store (STS) consisting of the memory in storage for a short amount of time, and long-term store (LTS) refers to the memory maintained for long periods of time \cite{andersen2007hippocampus, olton1979hippocampus}.

Supporting this view, our model mimics the prefrontal cortex using Quantum Spiking Neural Networks (QSNN) and the hippocampus using Quantum Long Short-Term Memory (QLSTM). The QSNN effectively filters out noisy and infrequent events while strengthening information with stronger space-time correlations. Subsequently, we access QLSTM to retrieve specific information, engage in processing and memorization phases, and transform it from short-term to long-term storage. Given the temporal nature of QLSTM, we preserve both temporal and spatial information throughout the learning process.
Additionally, In the realm of traditional neural networks, the temporal constants of neurons and synapses typically span time scales ranging from 1 to 100 milliseconds. However, challenges arise when dealing with problems that necessitate long-term associations exceeding the slowest neuron or synaptic time constant. Such challenges are prevalent in natural language processing and reinforcement learning, crucial for understanding human behavior and decision-making. This imposes a substantial burden on the learning process, where vanishing gradients impede the convergence of neural networks \cite{SNNTraining}.

To address these challenges, Long Short-Term Memory networks (LSTMs) and later, Gated Recurrent Units (GRUs) \cite{cho-etal-2014-learning}, introduced slow dynamics designed to overcome memory and vanishing gradient issues in traditional Recurrent Neural Networks (RNNs). For spiking neural networks, complementing fast neural dynamics with a variety of slower dynamics becomes essential to effectively tackle long-range association problems \cite{SNNTraining}.

Motivated by this, the exploration of the union between Quantum Spiking Neural Networks (QSNN) and Quantum Long Short-Term Memory (QLSTM) becomes compelling. The quantum nature of these architectures holds the promise of overcoming temporal limitations, offering diverse dynamics adaptable to a broad range of time scales. This innovative combination not only has the potential to address persistent challenges in long-term association learning but may also unlock new possibilities in real-time dynamic information processing, providing a significant boost to the efficiency and adaptability of neural networks across various applications.

In this manuscript, we will conduct a comparative study in the context of anomaly detection, specifically targeting fraud detection in bank account transactions. We will evaluate three different architectures, including Spiking Neural Networks (SNN) and Artificial Neural Networks (ANN), as well as quantum architectures such as Quantum Spiking Neural Networks (QSNN) and Quantum Neural Networks (QNN). Additionally, we will explore the synergy between these quantum architectures and Quantum Long Short-Term Memory (QLSTM). Our research will focus on determining which of these architectures exhibits optimal performance in a scenario characterized by highly imbalanced data and limited data availability.

\section{Background}\label{background}
For article self-completeness, this section provides a concise introduction to Spiking Neural Networks (SNNs), Long Short-Term Memory, and Quantum Neural Networks (QNNs), which constitute the primary topics covered in this work.

Firstly, in Section \ref{sec:snn}, we delve into the fundamentals of Spiking Neural Networks. This section offers an overview of the basic structure and operation of SNNs, alongside a discussion of Hebbian Theory.

Subsequently, in Section \ref{sec:lstm}, we explore the architecture and operation of Long Short-Term Memory networks.

Lastly, in Section \ref{sec:qnn}, we introduce Quantum Neural Networks, elucidating their key concepts and principles.

\subsection{Spiking Neural Networks}\label{sec:snn}
SNNs are inspired by the brain's communication scheme between neurons, similar to the encoding and maintenance of working memory or short-term memory in the prefrontal cortex, as well as the Hebbian plasticity principle.

\begin{figure*}[ht!]
\centering
\includegraphics[width=\textwidth]{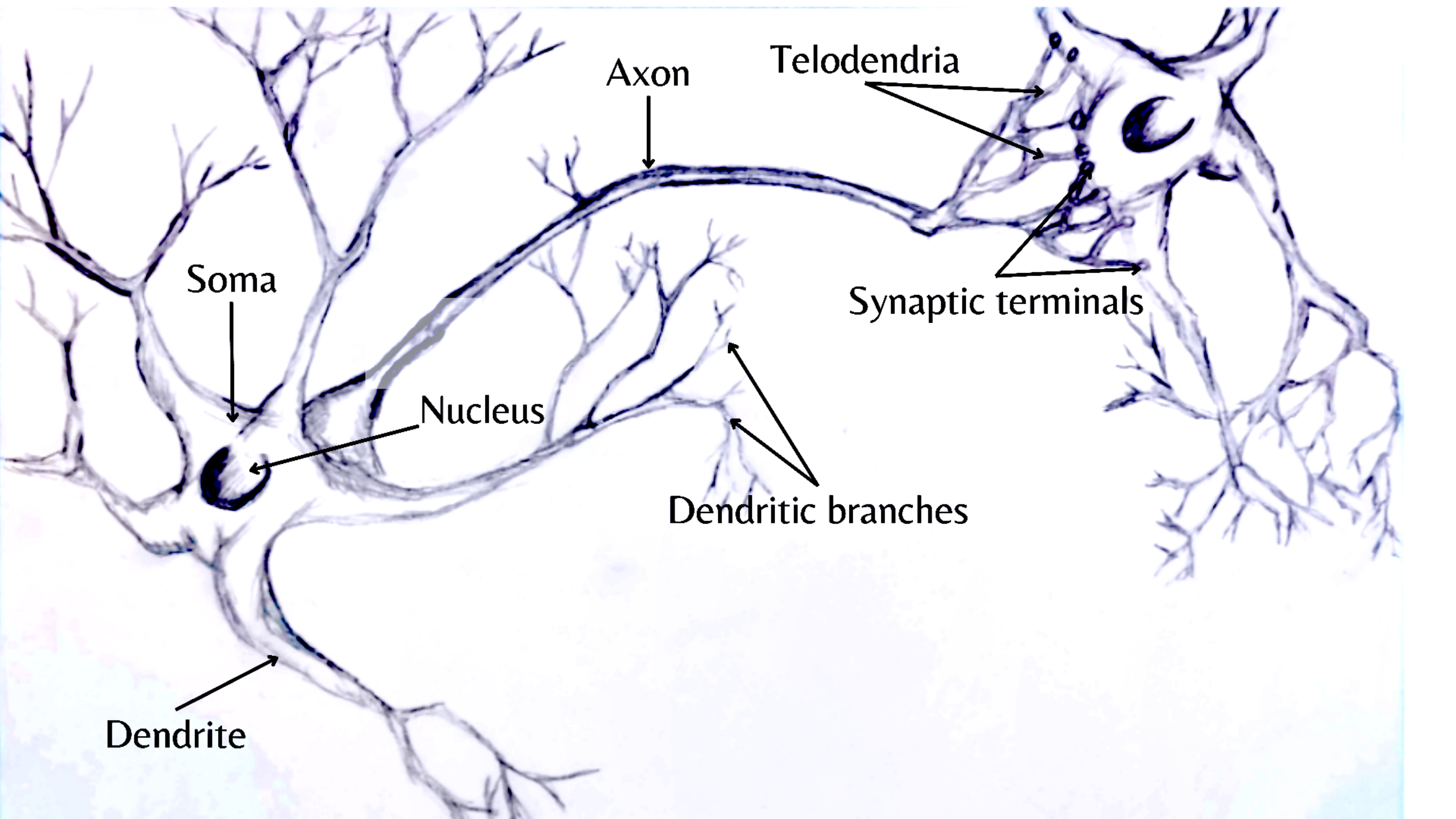}
\caption{Typical morphology of a neuron. Consisting of a cell body, or soma, which contains the nucleus and other organelles, dendrites, which are fine, branched cell processes that receive synaptic input from other neurons, one axon and synaptic terminals.}\label{fig:neuronmorphology}
\end{figure*}

The Hebbian theory is a neuroscientific concept that describes a fundamental mechanism of synaptic plasticity. According to this theory, the strength of a synaptic connection increases when neurons on both sides of the synapse are repeatedly activated simultaneously. Introduced by Donald Hebb in 1949, it is known by various names, including Hebb's rule, Hebbian learning postulate, or Cell Assembly Theory. The theory suggests that the persistence of repetitive activity or a signal tends to induce long-lasting cellular changes that enhance synaptic stability. When two cells or systems of cells are consistently active at the same time, they tend to become associated, facilitating each other's activity. This association leads to the development of synaptic terminals on the axon of the first cell in contact with the soma of the second cell, as depicted in Fig. \ref{fig:neuronmorphology} \cite{hebb2005organization}.

Although, DNNs are historically brain-inspired, there are fundamental differences in their structure, neural computations,and learning rule compared to the brain. One of the most important differences is the way that information propagates between their units. It is this observation that leads to the realm of spiking neural networks (SNNs). In the brain the communication between neurons is done by broadcasting trains of potentials or spike trains to downstream neurons. These individual spikes are sparse in time, so each spike has high information content. Thus, information in SNNs is conveyed by spike timing, including latencies, and spike rates. In a biological neuron, a spike is generated when the running sum of changes in the membrane potential, which can result from pre-synaptic stimulation, crosses a threshold. The rate of spike generation and the temporal pattern of spike trains carry information about external stimuli and ongoing calculations. 

ANNs communicate using continuous-valued activations and, for that reason, SNNs are more efficient because, as explained below, spike events are sparse in time. Additionally, SNNs also have the advantage of being intrinsically sensitive to the temporal characteristics of information transmission that occurs in the biological neural systems. The precise timing of every spike is highly reliable for several areas of the brain, suggesting an important role in neural codding (in sensing information-processing areas and neural-motor areas) \cite{SNNTraining, DLSNN}. 
SNNs have applications in many areas of pattern recognition such as visual processing, speech recognition and medical diagnosis. Deep SNNs are great candidates to investigate neural computation and different coding strategies in the brain. Training deep spiking neural networks is in its early phases and there is an important open question about their training such us let on-line learning while avoiding catastrophic forgetting \cite{SnnOL}.

\begin{figure*}
    \centering
    \begin{subfigure}{8cm}
    \includegraphics[width=\linewidth]{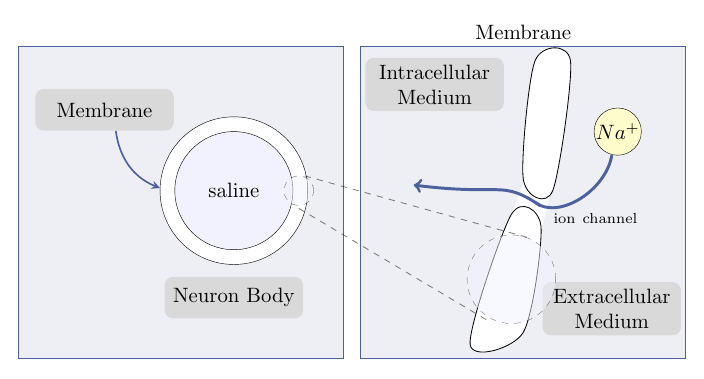}
    \caption{}\label{fig:b1}
    \end{subfigure}
    \qquad
    \begin{subfigure}{5cm}
    \includegraphics[width=\linewidth]{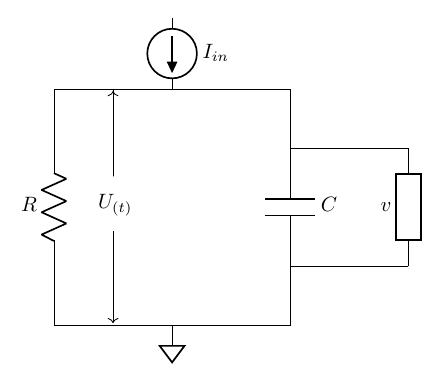}
    \caption{}\label{fig:b2}
    \end{subfigure}
    \begin{subfigure}{6cm}
    \includegraphics[width=\linewidth]{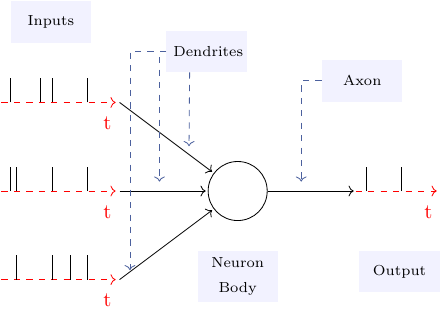}
    \caption{}\label{fig:b3}
    \end{subfigure}
    \begin{subfigure}{6cm}
    \includegraphics[width=\linewidth]{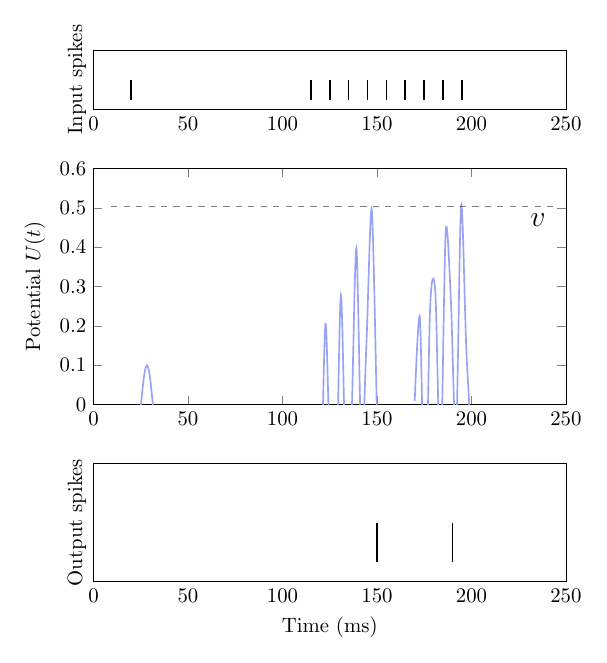}
    \caption{}\label{fig:b4}
    \end{subfigure}
    \caption{Leaky Integrate-and-Fire Neuron Model \cite{SNNTraining}. An insulating lipid bilayer membrane separates the interior and exterior environments. Gated ion channels enable the diffusion of charge carriers like Na+ across the membrane.\ref{fig:b1}. RC circuit models neuron function. When the membrane potential exceeds the threshold, a spike is generated. \ref{fig:b2}. Input spikes are transmitted to the neuron body through dendritic branches. Sufficient excitation accumulation triggers spike emission at the output \ref{fig:b3}. A simulation illustrating the membrane potential $U(t)$ with a threshold of $\theta = 0.5V$, resulting in the generation of output spikes \ref{fig:b4} }
    \label{fig:biologicalmodel}
\end{figure*}

Spiking neurons operate on a weighted sum of inputs. Rather than passing the result through a sigmoid or ReLU non-linearity, the weighted sum contributes to the membrane potential $U(t)$ of the neuron. If the neuron is sufficiently excited by this weighted sum, and the membrane potential reaches a threshold $\theta$, then the neuron will emit a spike to its subsequent connections. But most neuronal inputs are spikes of very short bursts of electrical activity. It is quite unlikely for all input spikes to arrive at the neuron body in unison. This indicates the presence of temporal dynamics that ‘sustain’ the membrane potential over time. Fig. \ref{fig:biologicalmodel}.

Louis Lapicque \cite{lapicque1907recherches} observed that a spiking neuron can be roughly compared to a low-pass filter circuit, consisting of a resistor (R) and a capacitor (C). This concept is known as the leaky integrate-and-fire neuron model. Even today, this idea remains valid. Physiologically, the neuron's capacitance is due to the insulating lipid bilayer forming its membrane, while the resistance comes from gated ion channels that regulate the flow of charged particles across the membrane (see Fig. \ref{fig:b2}).

The behavior of this passive membrane can be described using an RC circuit, following Ohm's Law. It states that the membrane's potential, measured between the input and output of the neuron, is proportional to the current passing through the conductor.
\begin{equation}
        I_{in}(t) = \frac {U(t)}{R}
    \end{equation}
     The dynamics of the passive membrane modelled using an RC circuit can be represented as:
    \begin{equation}
        \tau \frac{dU(t)}{dt} = - U(t) + I_{in}(t)R        
    \end{equation}
    where $\tau=RC$ is the time constant of the circuit.
    Following the Euler method, $dU(t)/dt$ without $\Delta t \to0$:
    \begin{equation}
        \tau \frac{U(t + \Delta t)-U(t)}{\Delta t} = - U(t) + I_{in}(t)R 
    \end{equation}
    Isolating the membrane potential at the next step:
    \begin{equation}
        U(t + \Delta t)= \left(1 - \frac{\Delta t}{\tau}\right)U(t) + \frac{\Delta t}{\tau} I_{in}(t) R
    \end{equation}
    To single out the leaky membrane potential dynamics, assume there is no input current $I_{in}= 0$:
    \begin{equation}
        U(t + \Delta t) = \left(1 - \frac{\Delta t}{\tau}\right)U(t)
    \end{equation}
$\beta=U(t + \Delta t)/U(t)$ would be the decay rate of the membrane potential, also known as the inverse time constant. From the previous equation, this implies that $\beta=1 - \Delta t/\tau$.

Let us assume that $t$ is discretized into sequential time-steps, such that $\Delta t = 1$. To further reduce the number of hyperparameters assume $R = 1\,\Omega$. Then

\begin{equation}
  \beta= 1 - \frac{1}{\tau} \Rightarrow U(t+1)= \beta U(t)+(1-\beta)I_{in}(t+1)
\end{equation}
For a constant current input, this can be solved as
     \begin{equation}
         U(t)= I_{in}(t) R + [U_0 - I_{in}(t) R]e^{-t/\tau}
     \end{equation}
which shows how exponential relaxation of $U(t)$ to a steady state value follows current injection, where $U_0$ is the initial membrane potential at $t = 0$
\begin{equation}\label{eq:u0ut}
    U(t) =U_0 e^{-t/\tau}
\end{equation}
Assuming Eq.~\eqref{eq:u0ut} is computed at discrete steps of $t, (t+\Delta t), (t + 2\Delta t)...,$ then the ratio of membrane potential across two subsequent steps can be calculated using:

\begin{equation} \label{eq:ratio}
    \beta = \frac{U_0e^{-(t+\Delta t)/\tau}}{U_0e^{-t/\tau}} = \frac{U_0e^{-(t+2\Delta t)/\tau}}{U_0e^{-(t+\Delta t)/\tau}}= ... \Rightarrow \beta = e^{-\Delta t/\tau}
\end{equation}
This equation for $\beta$ is more precise than $\beta = (1- \Delta t/\tau)$, as the latter is only precise for $\Delta t \ll \tau$.

A second non-physiological assumption is made, where the effect of $(1- \beta)$ is absorbed by a learnable weight $W$ (in deep learning, the weighting factor of an input is typically a learnable parameter):

\begin{equation}
   WX(t)= I_{in}(t)    
\end{equation}
$X(t)$ is an input voltage, spike, or unweighted current, and is scaled by the synaptic conductance $W$ to generate a current injection to the neuron. This leads to the following result:
\begin{equation}
    U(t+1)= \beta U(t) + WX(t+1)
\end{equation}
where the effects of $W$ and $\beta$ are decoupled, thus favouring simplicity over biological precision.
Finally, a reset function is appended, which activates every time an output spike is triggered:

\begin{equation}
    U(t)=  \underbrace{\beta U(t-1)}_\text{decay} +  \underbrace{WX(t)}_\text{input} -  \underbrace{S_{out}(t-1)\theta}_\text{reset}    
\end{equation}
where $S_{out}(t) \in \{0,1\}$ is the output spike, 1 in case of activation and 0 in otherwise. In the first case, the reset term subtracts the threshold $\theta$ from the membrane potential and in the second case, the reset term has no effect.

A spike is generated if the membrane potential exceeds the threshold:
\begin{equation}\label{eq:heavisite}
S_{out}(t) = 
\begin{cases}
      1, & \text{if}\ U(t) > \theta \\
      0, & \text{otherwise}
    \end{cases}\end{equation}

There are several methods to training SNN's \cite{SNNTraining}, the more commonly adopted approach is the backpropagation using spikes i.e backpropagation through time (BPTT). Working backwards from the final output of the network, the gradient flows from the loss to all descents.
The goal is to train the network using the gradient of the loss with respect to the weights, such that the weights are updated to minimize the loss. The backpropagation algorithm achieves this using the chain rule:
\begin{equation}
\frac{\partial \mathcal{L}}{\partial W} = 
\frac{\partial \mathcal{L}}{\partial S}
\underbrace{\frac{\partial S}{\partial U}}_{\{0, \infty\}}
\frac{\partial U}{\partial I}\
\frac{\partial I}{\partial W}
\end{equation}

However, the derivative of the Heaviside step function from \eqref{eq:heavisite} is the Dirac Delta function, which evaluates to 0 everywhere, except at the threshold $U_{\rm thr} = \theta$, where it tends to infinity. This means the gradient will almost always be nulled to zero (or saturated if $U$ sits precisely at the threshold), and no learning can take place. This is known as the dead neuron problem. The most common way to address the dead neuron problem is to keep the Heaviside function as it is during the forward pass, but during the backward, substitute the heaviside operator with a continuous function, $tilde(S)$ the derivative of the continuous function is used as a substitute $\partial S/\partial U \leftarrow\partial \tilde{S}/\partial U$, and is known as the surrogate gradient. In this manuscript we use snnTorch library which uses arctangent as a default function \cite{SNNTraining}.

\begin{figure*}
\centering
\includegraphics[width=\textwidth]{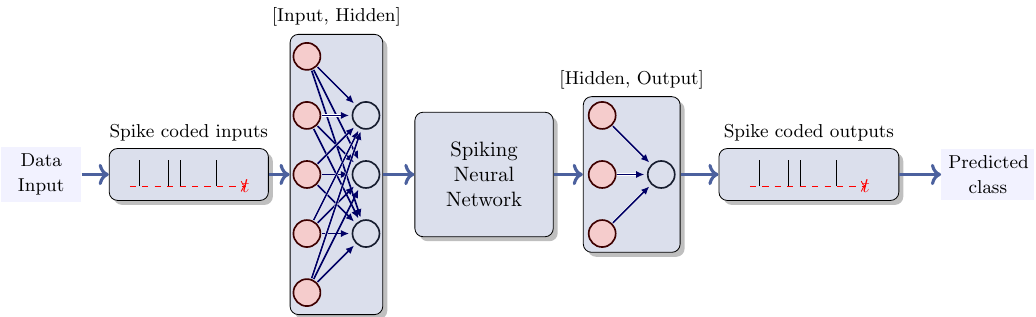}
\caption{\textbf{SNN pipeline.} Input data for an SNN can be transformed into a firing rate or other encodings to generate spikes. The network is subsequently trained to predict the correct class, employing encoding strategies such as the highest firing rate or firing first, among others}
\label{fig:snnscheme}
\end{figure*}

The structure of QSNNs follows a hybrid architecture formed by classical linear layers and a variational quantum circuit (VQC) for the implementation of the quantum leaky integrate-and-fire (QLIF) neuron, trained using the gradient descent method. Figure \ref{fig:snnscheme} shows the general pipeline for this model in a classification task, and the detailed architecture is defined in Section \ref{sec:methodology}.

Previous works have been made inspiring in brain functionality emulating for clasification tasks such as MINST dataset classication using SNN and Hyperdimensional computing \cite{SnnHdc} or decoding and understanding muscle activity and kinematics from electroencephalography signals \cite{snnmuscle}, using Hyperdimensional Computing and Spiking (HDC) and SNN for MNIST classification problem \cite{SnnHdc} or using Reinforcement Learning for navigation in changeable and unfamiliar environments supporting neuroscientific theories that see grid cells as critical for vector-based navigation \cite{navigationRL}.

\subsection{Long Short-Term Memory}\label{sec:lstm}
Long Short-Term Memory (LSTM) networks are a type of recurrent neural network capable of learning order dependence in sequence prediction problems. 
They are designed to overcome the RNN (Recurrent Neural Networks) problems in training because retro-propagated gradients tend to grow enormously or fade over time because the gradient depends not only on the present error but also on past errors. The accumulation of errors makes it difﬁcult to memorize long-term dependencies. Therefore, these problems are solved by the Long Short-Term Memory neural networks (LSTM), for which it incorporates a series of steps to decide which information will be stored and which erased \cite{lstmlongtermdependencies}.
Additionally, the range of contextual information that standard RNNs can access in practice is quite limited. The influence of a given input on the hidden layer and, therefore, on the network output, either decays or blows up exponentially as it cycles around the network's recurrent connections. This is the vanishing gradient problem, being the second problem to overcome using LSTM~\cite{lstmhandwritting, lstmclassification}.

The behavior of this model is required in complex problem domains like machine translation, speech recognition and time-series among others.

\begin{figure}[ht!]
    \centering
    \includegraphics[width=\columnwidth]{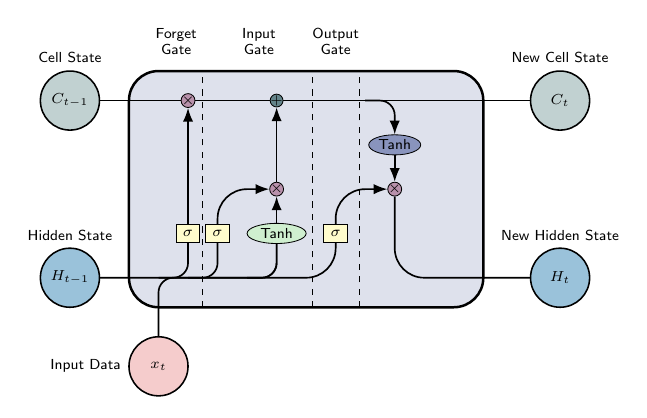}
    \caption{ \textbf{LSTM Cell Architecture:} Featuring three essential gates (forget, input and output gates). The $\sigma$ and $tanh$ blocks symbolize the sigmoid and hyperbolic tangent activation functions, respectively. $x_t$ denotes the input at time $t$, $h_t$ represents the hidden state, and $c_t$ signifies the cell state. The symbols $\otimes$ and $\oplus$ denote element-wise multiplication and addition, respectively.}
    \label{fig:lstmarch}
\end{figure}

These networks are composed of LSTM modules which are a special type of recurrent neural network described in 1997 by Hochreiter and Schmidhuber \cite{Lstmsolve}. They consist of three internal gates, known as input, forget and output gates, detailed in Fig. \ref{fig:lstmarch}.

These gates are filters and each of them have its own neural-network.
The output of an LSTM at a particular point in time is dependant on three things:
\begin{itemize}
    \item Cell state: The current long-term memory of the network 
    \item Hidden state: The output at the previous point in time 
    \item The input data at the current time step
\end{itemize}
The internal gates mentioned above can be described as follows \cite{lstm}:
\begin{itemize}
    \item \textbf{Forget Gate}: This gate determines which data from the cell state is relevant based on the previous hidden state and the new input data. The neural network that implements this gate is build to achieve an output closer to 0 when the input data is considered irrelevant, and closer to 1 otherwise. To achieve this, we employ the sigmoid activation function. The output values from this gate are then sent upward and subjected to pointwise multiplication with the previous cell state. This pointwise multiplication implies that components of the cell state deemed irrelevant by the forget gate network will be multiplied by a number close to 0, consequently having less influence on subsequent steps.

    In summary, the forget gate decides which pieces of the long-term memory should now be forgotten (have less weight) given the previous hidden state and the new input data.
    \item \textbf{Input gate}: Determines which new information should be incorporated into the network's long-term memory (cell state), based on the previous hidden state and new input data. The same inputs are utilize, but with the addition of a hyperbolic tangent as the activation function. This hiperbolic tangent has learned to combine the previous hidden state and new input data, resulting in a new memory update vector. Essentially, this vector contains information from the new input data within the context of the previous hidden state. This vector tells us how much to update each component of the long-term memory (cell state) of the network given the new data.

    Note that the utilization of the hyperbolic tangent function in this 
 context is deliberate, owing to its output range confined to [-1,1]. The inclusion of negative values is imperative for this methodology, as it facilitates the attenuation of the impact associated with specific components.

    \item \textbf{Output gate}: the objective of this gate is to decide the new hidden state by incorporating the newly updated cell state, the previous hidden state, and the new input data. This hidden state has to contain the necessary information while avoiding the inclusion of all learned data. To achieve this, we employ the sigmoid function.

\end{itemize}

This architecture is replicated for each time step considered in the prediction. The final layer of this model is a linear layer responsible for converting the hidden state into the ultimate prediction.

The quantum counterpart of this neural network is constructed with a VQC model for each gate as Fig. \ref{fig:QLSTMArch} shows.

Finally, we sumarise the Lstm implementation steps as follows:

\begin{itemize}
    \item The first step consists of deciding which information is to be forgotten or retained in the particular time instant, to achieve this, the sigmoid function is employed. It analyzes the previous state $h_{t-1}$ and the current input $x_t$, calculating the function accordingly:
    \begin{equation}
        f_t= \sigma(W_f \cdot v_t + bf)  
    \end{equation}
    Where  $v_t=[x_t, h_{t-1}]$ and $w_f$ and $b_f$ are weights and biases.

    \item  In this step, the content of the memory cell is updated by  selecting new information to be stored in the cell state. The second layer, known as the input gate, comprises two components: the sigmoid function and the hyperbolic tangent (tanh). The sigmoid layer determines which values are to be updated; a value of 1, indicates no change, while a value of 0 results in exclusion. Subsequently, a tanh layer generates a vector of new candidate values, assigning weights to the selected values based on their importance (ranging from (-1 to 1). These tow components are then combined to update the state:
    \begin{equation}
        \begin{split}
            i_t= \sigma(W_i \cdot v_t + b_i) \\
            \widetilde{C}= tanh(W_c \cdot v_t + b_c)
        \end{split}    
    \end{equation}

    \item The third step consists of updating the old cell state, $C_{t-1}$ with the new cell state, $C_t$, by multiplying the old state by $f_t$ to forget irrelevant information, and then adding the candidate $\widetilde {C_t}$:
     \begin{equation}
         C_t= f_t \cdot c_{t-1} + i_t \cdot \widetilde{C_t}
     \end{equation}

     \item Finally, the output is computed in two steps. Firstly, a sigmoid layer is used to select the relevant portions of the cell state to be transmitted to the output.
     \begin{equation}
         o_t= \sigma(W_o \cdot v_t + b_0)         
     \end{equation}
     The cell state is then processed through the tanh layer (to normalize values within the range of -1 and 1) and multiplied by the output of the sigmoid gate.
     \begin{equation}
         h_t= tanh(C_t) \cdot o_t
     \end{equation}
\end{itemize}

\subsection{Quantum Neural Networks}\label{sec:qnn}

Quantum Neural Networks (QNNs) play a pivotal role in Quantum Machine Learning (QML) \cite{VQA_for_QNN, QML}, utilizing Variational Quantum Circuits (VQCs). These VQCs are hybrid algorithms, combining quantum circuits with trainable parameters optimized using classical techniques. They are capable of approximating continuous functions \cite{QNN-layers, VQA_for_QNN}, enabling tasks such as optimization, approximation, and classification.
Recent studies have also explored the application of QNNs in reinforcement learning and generative learning. For instance, \cite{QRL-energies} investigates the integration of variational quantum circuits in reinforcement learning tasks. Similarly, \cite{QuantumGenerativeLearning} provides a comprehensive review of quantum generative learning models, including quantum circuit Born machines and quantum generative adversarial networks. Additionally, \cite{QuantumSelfAttention} introduces a self-attention mechanism into QNNs. These works highlight the potential of quantum neural networks in artificial intelligence.

\begin {figure*}[ht!]
\centering
\includegraphics[width=\textwidth]{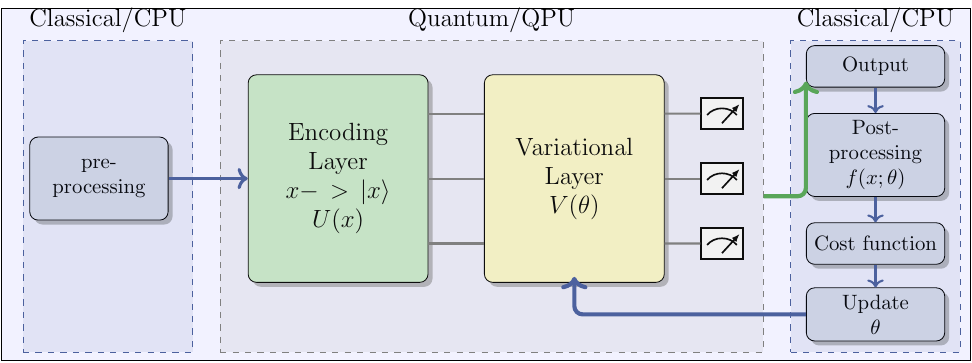}
\caption{\textbf{General VQC Schema.} The dashed gray line encompasses the steps executed in a Quantum Processing Unit (QPU) and the dashed blue line shows the steps executed in a CPU.}\label{fig:vqc-architecture}
\end{figure*}

Fig. \ref{fig:vqc-architecture} shows the general schema of a VQC. This hybrid methodology encompasses the following steps \cite{QMLBook}:

The VQC workflow comprises several steps:
\begin{enumerate}
\item \textbf{{Pre-processing (CPU)}}: Initial classical data preprocessing, which includes normalization and scaling.

\item \textbf{{Quantum Embedding (QPU)}}: Encoding classical data into quantum states through parameterized quantum gates. Various encoding techniques exist, including tensor product encoding and probability encoding  \cite{Expanding-DataEncoding}.

\item \textbf{{Variational Layer (QPU)}}: This layer implements the behavior of Quantum Neural Networks using entanglement and rotation gates with learnable parameters, optimized by classical optimization algorithms.

\item \textbf{{Measurement Process (QPU/CPU)}}: Measuring the quantum state and decoding it to obtain the desired output. The choice of observables used for this process is crucial for performance.

\item \textbf{{Post-processing (CPU)}}: Transformation of QPU outputs before returning them to the user and incorporating them into the cost function during the learning process.

\item \textbf{{Learning (CPU)}}: Computation of the cost function and optimization of ansatz parameters using classic optimization algorithms, such as Adam or SGD. Gradient-free methods like COLYBA or SPSA can also approximate parameter updates.
\end{enumerate}

\section{Methodology}\label{sec:methodology}

In this manuscript, we conducted an extensive benchmarking analysis, comparing various quantum-classical networks with their classical counterparts trained using gradient-descent. These include Quantum Neural Network (QNN), Quantum Spiking Neural Network (QSNN), and Quantum Long Short-Term Memory (QLSTM). Additionally, we introduce a novel model designed to closely emulate brain functionality by integrating QSNN and QLSTM, establishing a scalable and robust cognitive learning system. 
To achieve this, the QSNN component captures information at a low-level perspective, mirroring the role of the hypothalamus. Subsequently, the QLSTM processes this information at a higher level, identifying and memorizing correlated patterns while reinforcing long-term memorization, emulating hippocampus and mitigating the risk of catastrophic forgetting \cite{SNNTraining}. Catastrophic forgetting, where new information causes the network to forget what it has previously learned \cite{Catastrophic} is especially problematic in real-time systems where the size of batch-size is 1. Several approaches to overcome catastrophic forgetting in continual learning have been proposed, including using higher dimensional synapses \cite{continualLearning}, ensembles of networks \cite{ProgressiveNeuralNetworks}, pseudo-replay \cite{GenerativeReplay} and penalizing weights that change excessively fast \cite{overcomingcatastrhophic}. 
In summary, this innovative approach combines the advantages of QSNN for new learning with QLSTM's ability to retain and build upon previous knowledge. 

Considering the diminishing learning rate as network parameters approach optimal values, wherein future data has less influence than past data, our model proactively aims to prevent catastrophic forgetting by preserving previously acquired knowledge, adding a QLSTM model to the existing and trained QSNN. This proactive approach ensures a more comprehensive and stable learning system that balances the integration of new information with the retention of valuable past knowledge.

\begin{figure*}[ht!]
\centering
\includegraphics[width=\textwidth]{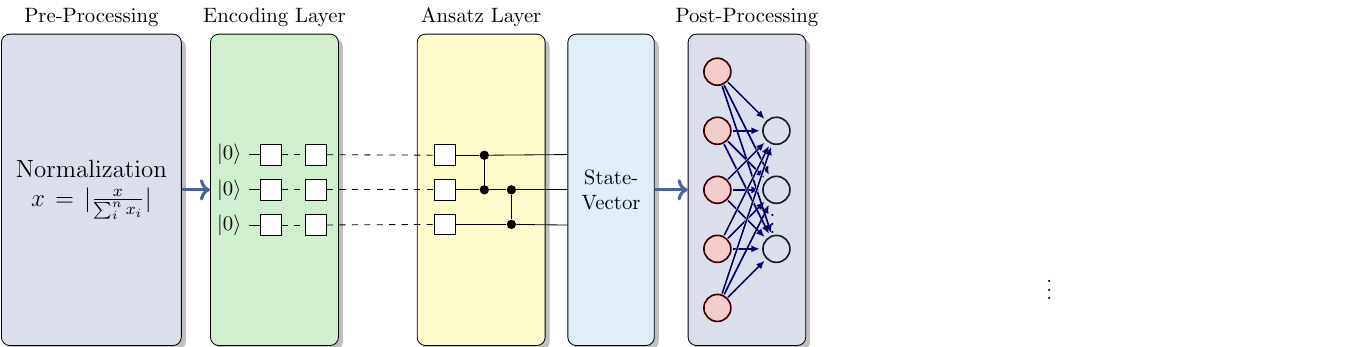}

\caption{\textbf{QNN architecture.}
This architecture entails a pre-processing step where classical data undergoes normalization, a pre-requisite for the subsequent encoding strategy. Then, the amplitud encoding algorithm is applied, utilizing only $log_2 N$ qubits, where N correspond to number of independent variables. This process generates the corresponding quantum state,  which is then feed into the ansatz circuit. Finally, the resulting state-vector of this quantum system is subject to post-processing through a classical linear layer. This layer transforms the dimension obtained with $2^{n}$, where n is the number of qubits, into the number of classes.}
\label{fig:QNNarch}
\end{figure*}

The initial model presented is a quantum neural network leveraging amplitude encoding to minimize qubits usage. This strategy is chosen because it necessitates only $log _2 n$ for n features. Fig. \ref{fig:QNNarch} illustrates this architecture.

\begin{figure*}[ht!]
\includegraphics[width=\textwidth]{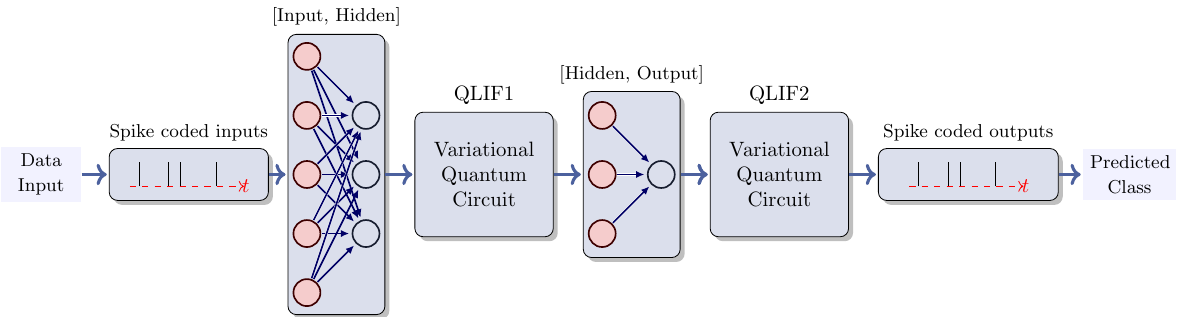}
\caption{\textbf{QSNN Architecture.}
This architecture involves an encoding step where classical data is translated into spikes. The network comprises two LIFs orchestrated by VQC and is trained using gradient descent to accurately predict the correct class using various encoding strategies, including the utilization of the highest firing rate or firing first, among others.}\label{fig:QSNNarch}
\end{figure*}

The second model is a Quantum Spiking Neural Network (QSNN), implemented using the Python package snnTorch \cite{snntorch}. The architecture of the QSNN is illustrated in Fig. \ref{fig:QSNNarch}. We enhanced the snnTorch library by incorporating a quantum version of a Leaky Integrate-and-Fire (LIF) neuron. In this quantum version, a VQC is employed to initialize the membrane potential, replacing the conventional tensor approach. The VQC consists of an encoding circuit using amplitude embedding to minimize qubit usage and an Ansatz composed of $R_x, R_y, R_z,$ and $C_z$ gates.

\begin{figure*}[ht!]
\centering
\includegraphics{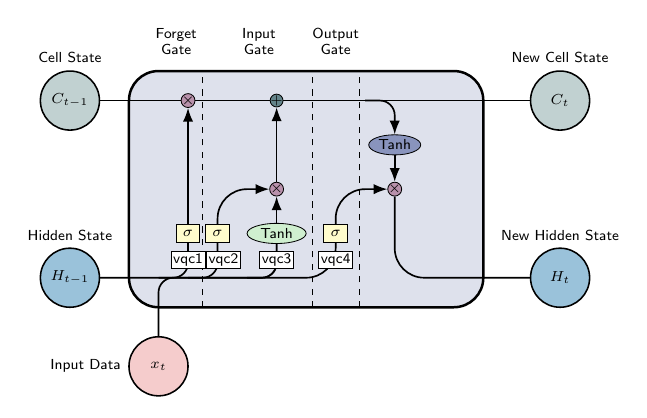}
\caption{\textbf{QLSTM Cell.} Each VQC box is the form as detailed in Fig. \ref{fig:vqc-architecture}. The $\sigma$ and $tanh$ blocks represent the sigmoid and the hyperbolic tanget activation function, respectively. $x_t$ is the input at time t, $h_t$ is the hidden state and $c_t$ is the cell state. $\otimes$ and $\oplus$ represents element-wise multiplication and addition, respectively.}\label{fig:QLSTMCell}
\end{figure*}

\begin{figure*}[ht!]
\centering
\includegraphics[width=\textwidth]{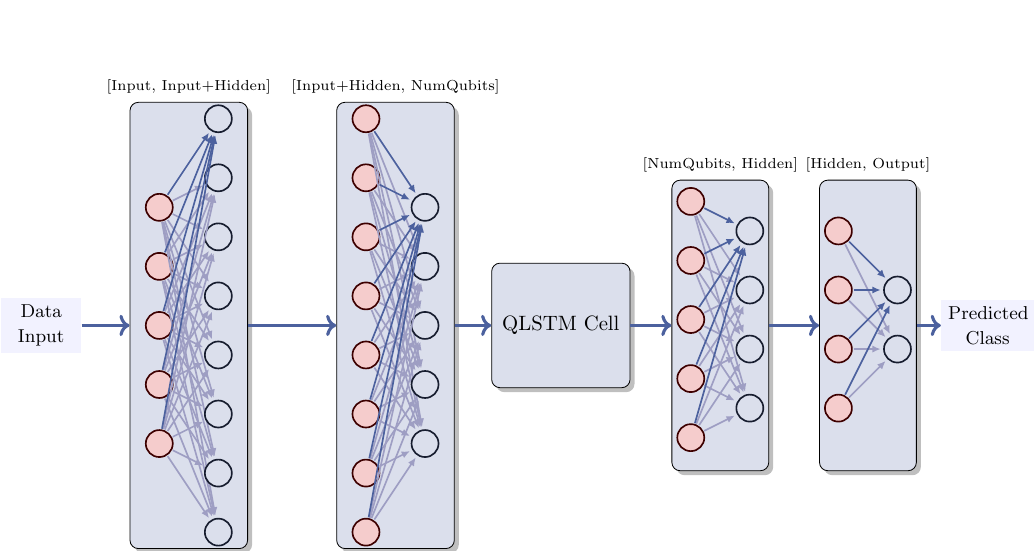}
\caption{\textbf{QLSTM Architecture.} The input data passes trough an initial classical layer, which receives the 30 inputs data and produces 80 outputs, corresponding to the concatenated size formed by the input dimension and hidden size. This output then passes trough a second classical layer, which outputs the same size as the number of qubits expected by the Quantum QLSTM cell, whose architecture is detailed in Fig. \ref{fig:vqc-architecture}. Subsenquently, this output is received by another classical layer that transform it into output of the hidden size. Finally, this output is futher transformed into the number of classes, and by applying sigmoid activation function, we obtain the expected output $y_pred$.}\label{fig:QLSTMArch}
\end{figure*}

The third model is a Quantum Long Short-Term Memory (QLSTM) that utilizes VQCs for its forget, input, update, and output gates, the Quantum LSTM cell is detailed in Fig. \ref{fig:QLSTMCell} and the complete architecture in Fig. \ref{fig:QLSTMArch}. The encoding circuit employs Angle encoding after a Classical linear layer to transform concat size (input and hidden dimension) to number of qubits. Additionally, the ansatz circuits incorporate Basic Entangling Layers, which uses $R_x$ and $C_x$ gates.

\begin{figure*}
    \centering
    \begin{subfigure}{15cm}
    \includegraphics[width=\textwidth]{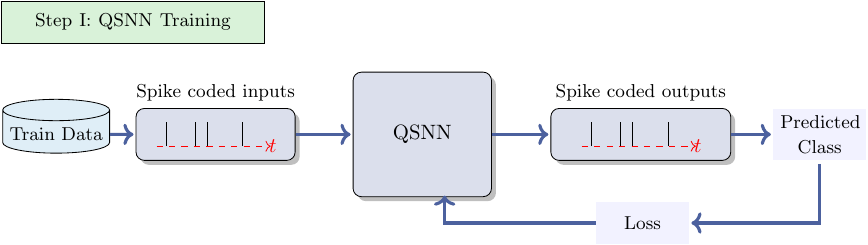}
    \caption{QSNN Training.}\label{fig:QSnnQLstm1}
    \end{subfigure}
    \qquad
    \begin{subfigure}{10cm}
    \centering
    \includegraphics[width=\textwidth]{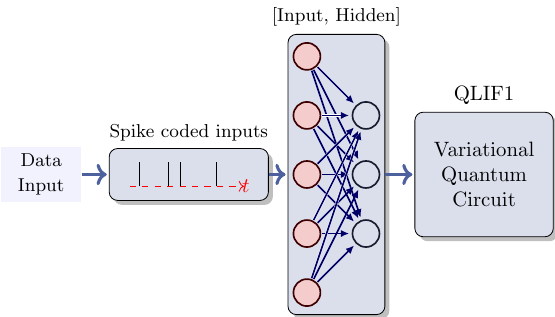}
    \caption{Split QSNN model.}
    \label{fig:qsnnqarchsplit}
    \end{subfigure}
    \begin{subfigure}{15cm}
    \includegraphics[width=\textwidth]{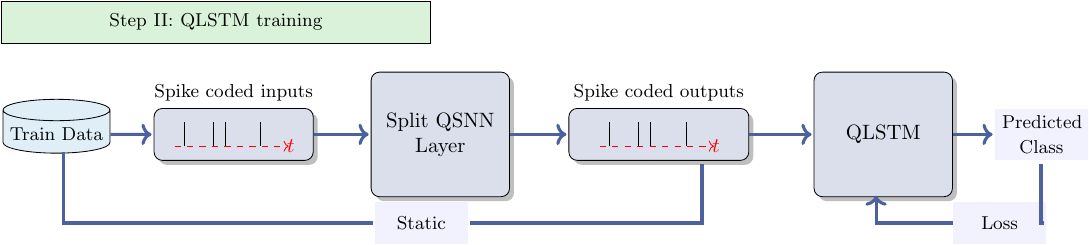}
    \caption{QLSTM training.}
    \label{fig:QsnnQlstm2}
    \end{subfigure}
    \begin{subfigure}{15cm}
    \includegraphics[width=\textwidth]{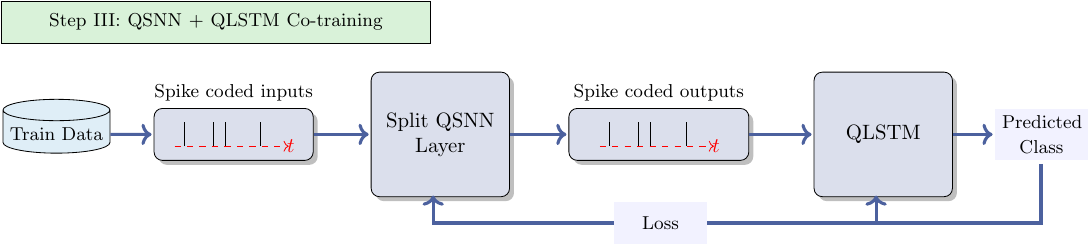}
    \caption{QSNN-QLSTM co-training.}
    \label{fig:QsnnQlstm3}
    \end{subfigure}
    \caption{\textbf{Training Process of QSNN-QLSTM.} Step I involves the iterative training of QSNN. An original instance of QSNN (see Fig. \ref{fig:QSNNarch}) is trained independently without the influence of QLSTM, utilizing the gradient descent method. The model is then split (see Fig. \ref{fig:qsnnqarchsplit}). In Step II, the split QSNN model is combined with a new instance of the QLSTM model for a single-pass training.During this step, the training data propagates from QSNN input layer through the QLSTM module, leading to modifications in the QLSTM's memory component. Step III Co-training of QSNN and QLSTM. The loss calculated from the QLSTM output is used to simultaneously update both models, adjusting them jointly to optimize the performance of the complete system.}\label{fig:QsnnQlstmCotraining}
\end{figure*}

The fourth quantum model is an innovative architecture that combines a Quantum Spiking Neural Network (QSNN) and Quantum Long Short-Term Memory (QLSTM). The training process involves three phases:
\begin{itemize}
    
\item Pre-training the QSNN: Initially, the QSNN is trained independently to adjust its parameters without the influence of the QLSTM.
\item Single-pass training of the QLSTM: The pre-trained QSNN is then integrated with a new instance of the QLSTM. In this phase, training data is propagated from the QSNN's input layer through to the QLSTM module, updating the QLSTM's memory components while keeping the QSNN layers static to ensure memory adaptation before co-training.
\item Co-training of QSNN and QLSTM: Finally, both models are trained jointly, with gradients updated simultaneously using a multi-optimizer that assigns distinct learning rates to each network's parameters.
\end{itemize}
This structured approach ensures cohesive learning and optimization of the combined QSNN-QLSTM architecture. The training process is illustrated in Fig. \ref{fig:QsnnQlstmCotraining}.

\section{Experimentation}\label{sec:experimentation}

In this section, we conduct experiments to evaluate various models described in the previous section for anomaly detection, with a specific focus on fraud detection in credit card transactions. Our objective is to compare the performance of all proposed models and determine whether the novel model composed of QSNN and QLSTM achieves superior results.
We performed 10 independent runs, each initiated with distinct random seeds, which were also used for dataset splitting. This approach ensures a diverse assessment across different partitions for each model.

Additionally, we utilize F1-score as the fundamental metric for the analysis and comparison of all models. 

All implemented models make use of the PyTorch, PennyLane and snnTorch libraries. 

\subsection{Problem Statement}

For the detection of fraud in credit card transactions, we utilize a Kaggle dataset comprising 30 features, which include 28 anonymized features resulting from PCA transformation and the transaction amount. The dataset, documenting transactions made by European cardholders in September 2013, spans a two-day period and includes a total of 284,807 transactions. Among these, 492 are identified as frauds. It is noteworthy that the dataset exhibits a pronounced class imbalance, with the positive class (frauds) representing a mere $0.172\%$ of the overall transactions.

\subsection{Experimental Settings}

For each execution, we recorded the AUC and the rates of true positives, true negatives, false positives, and false negatives for the calculation of F1-score, recall, and precision, as well as the training and validation losses, enabling a comprehensive analysis of their respective curves. Tables \ref{tab:ClassicalModelsConfiguration} and \ref{tab:QuantumModelsConfiguration} outline the configuration of classical and quantum models, along with the corresponding hyperparameters used.

For classical models, we utilized a larger dataset consisting of 390 fraud cases paired with 5000 non-fraud cases for training, and 999 instances of non-fraud paired with 101 instances of fraud for test dataset. In the case of quantum models, we utilised 390 fraud cases along with 1000 non-fraud cases for training dataset, mirroring the size of classical model's test dataset. This highlights that quantum models require fewer data points to achieve favorable metrics compared to classical models, showcasing their ability to discern non-trivial patterns that classical models struggle to identify, necessitating more data for effective learning. Additionally, We allocated 20 \% of training data for validation.

The first model, referred to as the QNN model, consists of the following components:
\begin{itemize}
    \item \textbf{Encoding and ansatz circuits:} The model is built with 5 qubits.
    \item \textbf{Number of qubits:} Determined by \( \log_2 N \), where \( N \) is the number of input features.
    \item \textbf{Linear layer:} Includes a layer that transforms \( 2^N \) into a single output.
    \item \textbf{Total parameters:}  
        \[
        \begin{aligned}
            &5 \times 3 \times 5 && (\text{layers} \times \text{rotations} \times \text{qubits}) \\
            &+ 2^5 (\text{statevector}) \times 1 + 1 && (\text{output linear layer}) \\
            &= 108.
        \end{aligned}
        \]
\end{itemize}    
The classical counterpart consists of the following layers:
\begin{itemize}
    \item \textbf{Input layer:} 30 neurons.
    \item \textbf{Hidden layers:}  
    \begin{itemize}
        \item First hidden layer: 20 neurons.  
        \item Second hidden layer: 5 neurons.  
    \end{itemize}
    \item \textbf{Output layer:} 1 neuron.
        \item \textbf{Total parameters:}  
        \[
        \begin{aligned}
            &30 \times 20, (\text{weights}) + 20 \, (\text{bias})  \\
            &+ 20 \times 5 \, (\text{weights}) + 5 \, (\text{bias}) \\
            &+ 5 \times 1 \, (\text{weights}) + 1 \, (\text{bias}) \\
            &= 731.
        \end{aligned}
        \]
\end{itemize}

The second model, QLSTM, as described in section \ref{sec:methodology}, consists of 6521 parameters and is structured as follows:
\begin{itemize}
    \item \textbf{A linear layer} that maps the input size to the input + hidden size:  
    \[
    30 \times (30 + 125) \, (\text{weights}) + (30 + 125) \, (\text{bias})
    \]
    \item \textbf{A linear layer} transforming the input and hidden size into 5 qubits:  
    \[
    (30 + 125) \times 5 \, (\text{weights}) + 5 \, (\text{bias})
    \]
    \item \textbf{The Variational Quantum Circuit (VQC)}, comprising 4 blocks, each with 3 layers, 5 qubits, and 1 trainable parameter used in the \( R_x \) gates of the ansatz circuit.
    \[
    4 \times 3 \times 5 \times 1 
    \]
    \item \textbf{A linear layer} mapping the qubit space back to the hidden size:  
    \[
    5 \times 125 \, (\text{weights}) + 125 \, (\text{bias})
    \]
    \item \textbf{A final linear layer} that transforms the hidden size into the output dimension:  
    \[
    125 \times 1 \, (\text{weights}) + 1 \, (\text{bias})
    \]
\end{itemize}

Its classical counterpart consists of:
\begin{itemize}
    \item A linear layer with 50 input neurons and 128 output neurons.
    \item Two LSTM units.
    \item A final layer with 128 input neurons and 1 output neuron.
    \item \textbf{Total parameters:}  
        \[
        \begin{aligned}
            & 4 \times 50 \times 30  \, (\text{input-hidden weights L0}) \\
            &+ 4 \times 50 \times 50  \, (\text{hidden-hidden weights L0})\\
            &+ 4 \times 50 \times 30 \, (\text{input-hidden bias L0})  \\
            &+ 4 \times 50 \times 30 \, (\text{hidden-hidden bias L0})  \\
            &+ 4 \times 50 \times 50  \, (\text{input-hidden weights L1})  \\
            &+ 4 \times 50 \times 50  \, (\text{hidden-hidden weights L1})\\
            &+ 4 \times 50 \times 30 \, (\text{input-hidden bias L1})  \\
            &+ 4 \times 50 \times 30 \, (\text{hidden-hidden bias L1})  \\
            &+ 50 \times 128 + 128 \, (\text{linear layer})  \\
            &+ 128 \times 1 + 1 \, (\text{final linear layer})  \\
            &= 43,457.
        \end{aligned}
        \]
\end{itemize}

The third model, QSNN described in section \ref{sec:methodology}, consists of the following components:
\begin{itemize}
    \item \textbf{QLIF cells:} The model has 2 QLIF cells.
    \item \textbf{Layers:} Each QLIF cell consists of 1 layer.
    \item \textbf{Hidden neurons:} 10 neurons in the hidden layer.
    \item \textbf{Qubits:} 5 qubits.
    \item \textbf{Total parameters:}  
        \[
        \begin{aligned}
            &30 \times 10  + 10 \, (\text{linear layer}) \\
            &+ 5 \times 3 \, (\text{3 rotation gates}) \times 2, (\text{QLifs}) \\
            &+ 10 \times 2 + 2 \, (\text{final layer})   \\
            &= 362.
        \end{aligned}
        \]
   
\end{itemize}

The classical counterpart of the QSNN model consists of the following elements:
\begin{itemize}
    \item \textbf{Linear layer:} 30 input neurons and 100 output neurons (hidden layer).
    \item \textbf{LIF neurons:} Two LIF neurons.
    \item \textbf{Final linear layer:} 100 input neurons and 2 output neurons (corresponding to the number of classes).
    \item \textbf{Total parameters:}  
        \[
        \begin{aligned}
            &30 \times 100 \, (\text{weights})  \\
            &+ 100 \, (\text{bias})  \\
            &+ 100 \times 2 , (\text{weights})  \\
            &+ 2 \, (\text{bias})  \\
            &= 3,302.
        \end{aligned}
        \]
    
\end{itemize}

The fourth model, QSNN-QLSTM, as described in Section \ref{sec:methodology}, consists of the following modules:
\begin{itemize}
    \item \textbf{QLIF cell:} 1 QLIF cell with 1 layer, 3 rotation gates,  10 neurons for the hidden layer, and 5 qubits.
    \item \textbf{QLSTM module:} 1 QLSTM module with 1 layer and 10 neurons for the hidden layer.
    \item \textbf{Total parameters:}  
        \[
        \begin{aligned}
            &30 \times 10 + 10 \, (\text{linear layer}) \\
            &+ 1 \times 5 \times 3 \, (\text{QLIF1})   \\
            &+ 10 \times 2 + 2 \, (\text{linear layer})   \\
            &+ 10 \times 20 + 20 \, (\text{QLSMT layer preprocessing}) \\
            &+ 20 \times 5 + 5  \, (\text{QLSTM layer in})\\
            &+ 5 \times 10 + 10 (\text{QLSTM layer out}) \\
            &+ 4\times 5 \times 1  (\text{QLSTM VQCs}) \\
            &+ 10 \times 2 + 2 (\text{final linear layer}) \\
            &= 774.
        \end{aligned}
        \] 

\end{itemize}

Our results show that quantum models generally require fewer parameters than their classical counterparts. When ordering the models from the fewest to the most parameters, we obtain the following sequence: QNN, QSNN, ANN, QSNN-QLSTM, SNN, QLSTM, LSTM. This highlights that even hybrid architectures like QSNN-QLSTM remain competitive, ranking fourth and closely following the ANN model despite their increased structural complexity.

\begin{table*}[ht!]
\centering
\fontsize{10pt}{10pt}\selectfont
\caption{Classical Models Configuration. Hyperparameters and configuration for Artificial Neural Network, Long Short-Term Memory and Spiking Neural Network respectively.}
\label{tab:ClassicalModelsConfiguration}
\begin{tabular}{|c|c|c|c|}
\hline
\textbf{} &\textbf{ANN} & \textbf{LSTM} & \textbf{SNN}\\ \hline
\textbf{Optimizer} & {SGD lr=1e-2} & {Adam lr=1e-3} & {Adam lr=1e-3}\\ \hline
\textbf{Scheduler} & {-} & {ReduceLROnPlateau} & {-}\\ 
\textbf{} & {} & {patience=30} & {}\\ \hline
\textbf{Batch Size} & {128} & {128} & {64}\\ \hline
\textbf{Iterations} & {700} & {350} & {350}\\ \hline
\textbf{Pre-processing} & {StandardScaler} & {StandardScaler} & {StandardScaler}\\ \hline
\textbf{Steps} & {-} & {-} & {25}\\ \hline
\textbf{Layers} & {[30, 20], Relu} & {[LSTM, 30, 50, layers=2]} & {[30 ,100]}\\ 
\textbf{} & {[20, 5], Relu} & {[50, 128]} & {Lif1 and Lif2}\\ 
\textbf{} & {[5, 1], Sigmoid}  & {[128, 1], Sigmoid } & {[100, 2]}\\ \hline
\end{tabular}
\end{table*}


\begin{table*}[ht!]
\centering
\fontsize{10.0pt}{10.0pt}\selectfont
\caption{Quantum Models Configuration. Hyperparameters and configuration for Quantum Neural Network, Quantum Long Short-Term Memory, Quantum Spiking Neural Network and the novel model composed by the union of last two networks.}

\label{tab:QuantumModelsConfiguration}
\begin{tabular}{|c|c|c|c|c|}
\hline
\textbf{} & \textbf{QNN} & \textbf{QLSTM} & \textbf{QSNN} & \textbf{QSNN-QLSTM}\\ \hline
\textbf{Optimizer} & {RMSprop lr=1e-2} & {Adam lr=1e-2} & {SGD lr=1e-3} & {Adam lr=1e-2}\\ 
\textbf{} & {} & {} & {} & {RMSprop lr=1e-2}\\ \hline

\textbf{Scheduler} & {-} & {RLROnPlateau} & {-} & {-}\\ 
\textbf{} & {} & {patience= 20} & {} & {}\\ \hline
\textbf{Batch Size} & {256} & {256} & {64} & {128}\\ \hline
\textbf{Iterations} & {70} & {100} & {80} & {40}\\ \hline
\textbf{Pre-processing} & {StandardScaler} & {MinMaxScaler(0,$\pi$/2)} & {StandardScaler} & {StandardScaler}\\ 
\textbf{} & {Normalization} & {} & {Normalization} & {Normalization}\\ \hline
\textbf{Post-processing} & {Sigmoid} & {Sigmoid} & {-} & {Sigmoid}\\ \hline
\textbf{Steps} & {-} & {-} & {25} & {25}\\ \hline
\textbf{Layers} & {2 QNN} & {Linear[125+30, 5]} & {Linear[30, 10]} & {1 QSNN}\\ 
\textbf{} & {Linear[$2^N$, 1]} & {1 QLSTM} & {1 QSNN} & {+}\\ 
\textbf{} & {} & {Linear[5, 125]} & {Linear[10, 2]} & {1 QLSTM}\\ 
\textbf{} & {} & {Linear[125, 1]} & {} & {}\\ \hline
\textbf{Qubits} & {5} & {5} & {5} & {5}\\ \hline
\textbf{Encoding} & {Amplitud} & {Angle} & {Amplitud} & {Amplitud}\\ 
\textbf{Strategy} & {Encoding} & {Encoding} & {Encoding} & {Encoding}\\ \hline

\end{tabular}
\end{table*}

The selection of hyperparameters, optimizers, and the number of iterations were conducted experimentally through multiple trials to achieve the best performance for each model. Specifically, the optimal number of iterations was determined by observing the point at which each model converged, thereby reducing training times. Early stopping was not employed to ensure that the ten runs of each model had a consistent number of iterations, facilitating a fair comparison.

\subsection{Results}

\begin{figure*}[ht!]
        \centering
        \includegraphics[width=\linewidth]{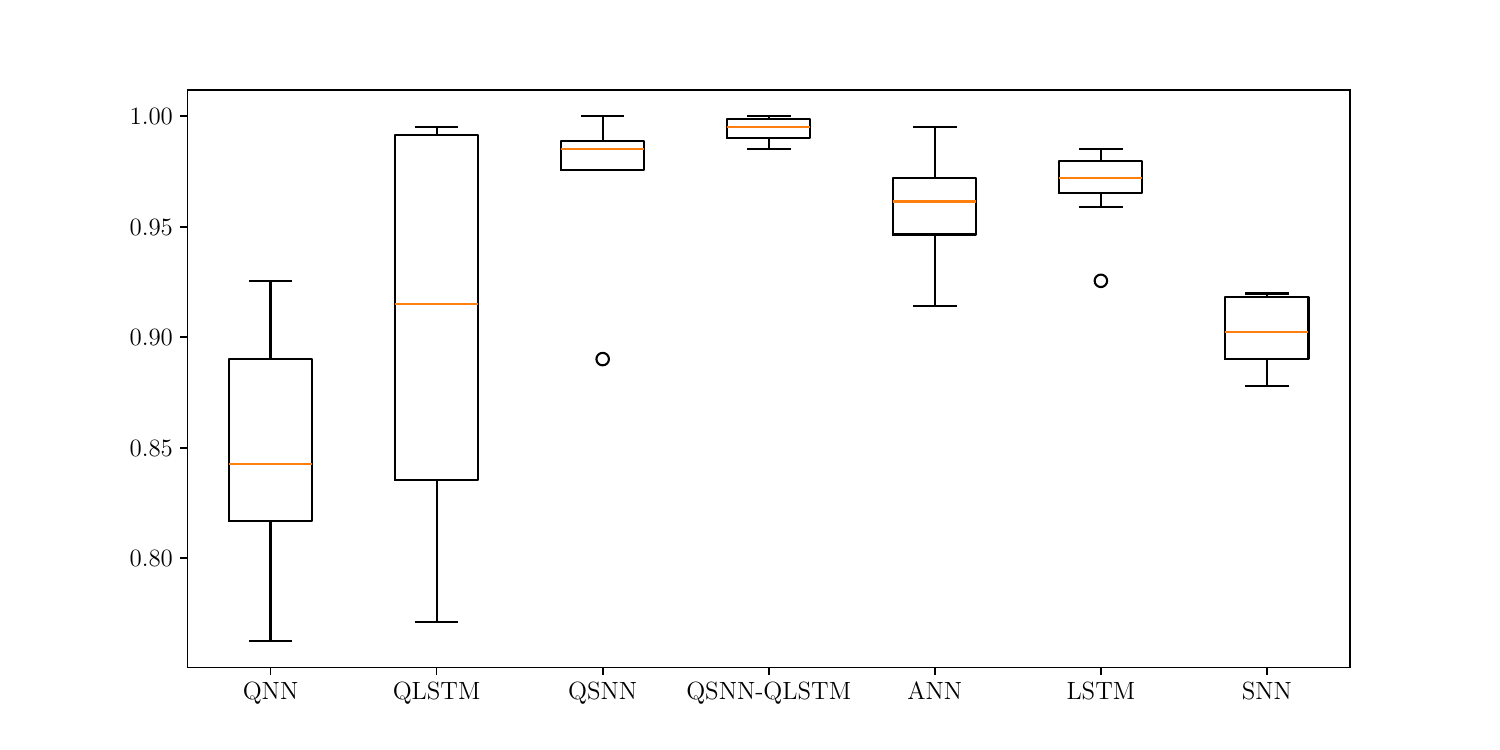}
\caption{Boxplots illustrating the distribution of F1 score obtained from quantum (QNN, QLSTM, QSNN and QSNN-QLSTM) and from classical models (ANN, LSTM and SNN).}
\label{fig:Boxplotf1}
\end{figure*}

\begin{figure*}[ht!]
        \centering
            \includegraphics[width=\linewidth]{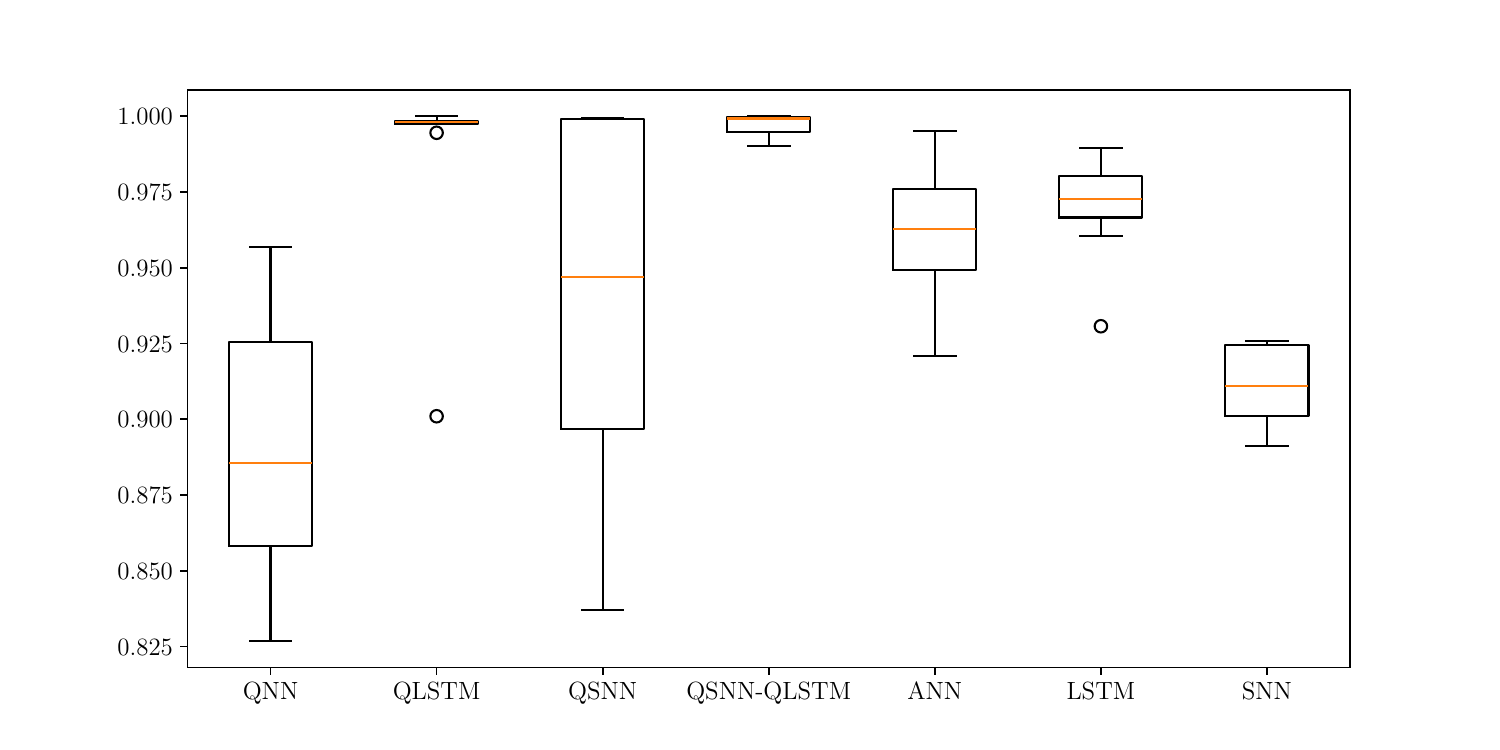}
\caption{Boxplots illustrating the distribution of AUC score obtained from quantum (QNN, QLSTM, QSNN and QSNN-QLSTM) and from classical models (ANN, LSTM and SNN).}
\label{fig:Boxplotauc}
\end{figure*}

\begin{figure*}[ht!]
        \centering
            \includegraphics[width=\linewidth]{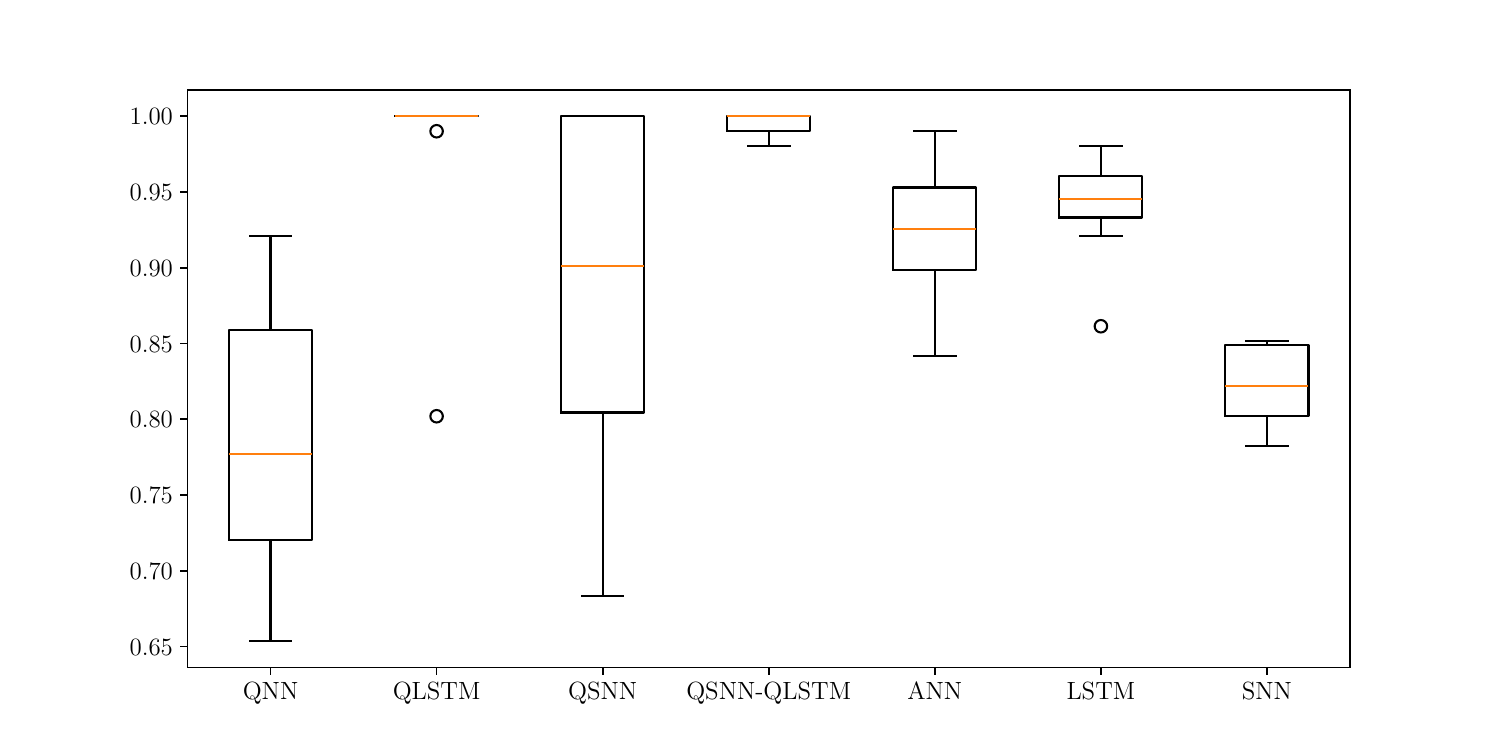}
\caption{Boxplots illustrating the distribution of Recall score obtained from quantum (QNN, QLSTM, QSNN and QSNN-QLSTM) and from classical models (ANN, LSTM and SNN).}
\label{fig:Boxplotrecall}
\end{figure*}

\begin{figure*}[ht!]
        \centering
            \includegraphics[width=\linewidth]{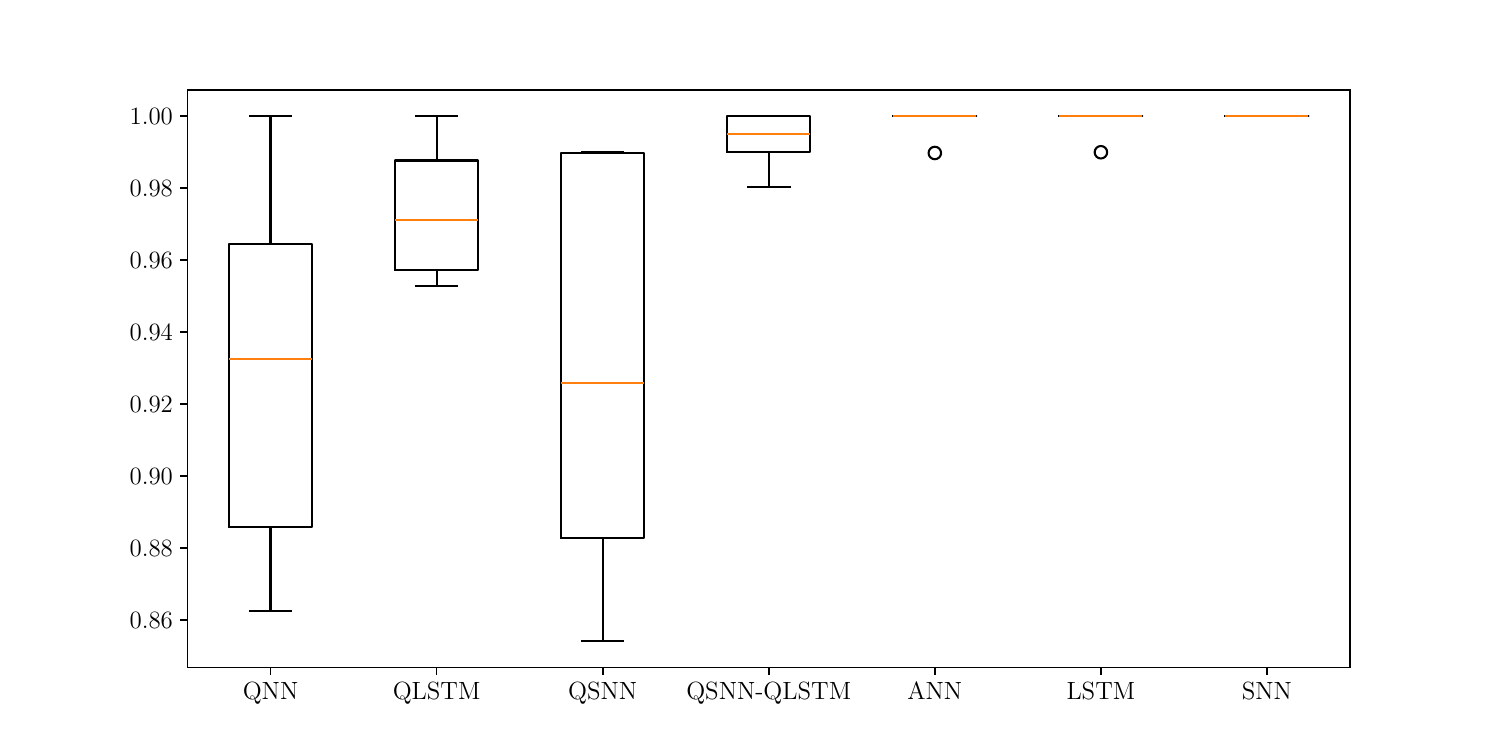}
\caption{Boxplots illustrating the distribution of Presicion score obtained from quantum (QNN, QLSTM, QSNN and QSNN-QLSTM) and from classical models (ANN, LSTM and SNN).}
\label{fig:Boxplotprecision}
\end{figure*}

The performance outcomes for this use case are depicted using boxplots: Fig. \ref{fig:Boxplotf1} illustrates the F1 score; Fig. \ref{fig:Boxplotauc}, the AUC score; Fig. \ref{fig:Boxplotrecall}, the recall score; and Fig. \ref{fig:Boxplotprecision}, the precision score.

Fig. \ref{fig:Boxplotf1} reveals that the QSNN-QLSTM model exhibits superior performance among quantum models, attaining the highest F1 score with notable robustness and requiring fewer iterations—a distinct advantage. The other models also display commendable robustness, with only a single outlier detected in the QSNN and LSTM models.

As depicted in Fig. \ref{fig:Boxplotauc}, the QSNN-QLSTM model exhibits the best performance and robustness in terms of AUC score. It is followed by the ANN, SNN and QNN models. In contrast, the QLSTM model exhibits the poorest robustness, attributed to the presence of two outliers.

Fig. \ref{fig:Boxplotrecall} illustrates that the QSNN-QLSTM model achieves the highest recall score and robustness, followed by the ANN, QNN, and SNN models.

Finally, Fig. \ref{fig:Boxplotprecision} shows that the SNN model achieves the highest precision and robustness, closely followed by the QSNN-QLSTM model. The remaining models also exhibit commendable performance and robustness.

In summary, across all boxplots, the QSNN-QLSTM consistently surpasses other models, demonstrating superior capability in capturing complex data relationships. This underscores the efficacy of QSNN-QLSTM as a neuroscience-inspired approach.

\begin{figure*}[htbp]
        \centering
            \includegraphics[width=\textwidth]{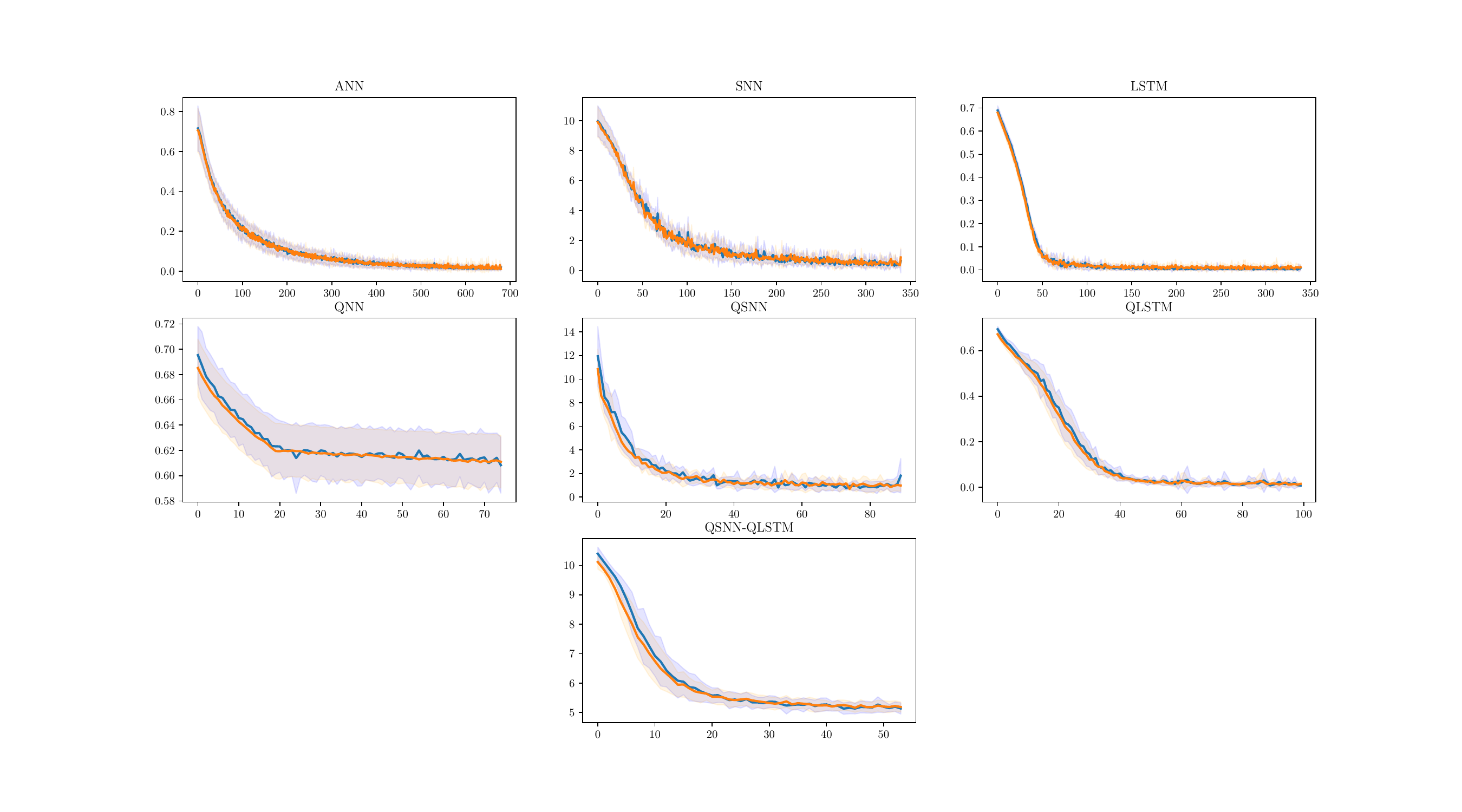}
\caption{Learning curves obtained from seven models. Three classical models: ANN, SNN and LSTM. Four quantum models: QNN, QSNN, QLSTM and new brain-inpired model QSNN-QLSTM.}
\label{fig:LearningCurves}
\end{figure*}

An additional advantage, beyond the reduced number of parameters and training data required by quantum models, is the fewer iterations needed, as observed in Fig. \ref{fig:LearningCurves}. This figure indicates that the QSNN-QLSTM model requires only 40 iterations, compared to the 700 needed by the ANN. However, it is important to note that the training time for quantum models is longer because simulating quantum operations on classical computers is highly resource intensive.

The remaining models exhibit strong robustness and performance in all metrics, demonstrating that the selected hyperparameters and architectures are appropriate.

Additionally, the learning curves in the same figure display a consistent progression across all models, with similar trends. Notably, variance decreases in later iterations, indicating increased stability and convergence. This suggests that the selected architectures effectively generalize over the dataset.

\section{Discussion}\label{sec:discussion}

We introduced three classical architectures and four quantum architectures, one of which incorporates a novel approach inspired by brain function to address the use case of fraud detection. A general perception derived from the analysis of the results is the effectiveness of utilizing quantum models inspired by the brain, achieving superior performance compared to classical approaches, while also exhibiting reduced complexity (requiring fewer parameters during training). Furthermore, quantum models were trained using a smaller amount of data, which presents a significant advantage. This not only contributes to reduced training times—particularly relevant when simulations are computationally expensive—but also offers clear benefits in real-world scenarios where data is scarce. Unlike classical approaches that often require synthetic data generation to augment training sets, quantum models may provide effective learning outcomes without the need for such techniques, which is a common challenge in data science.

On the other hand, although this model requires fewer iterations, as they have been executed in simulators, the runtime is impacted due to the computational load it carries. Accessing and conducting large-scale experiments on current quantum devices pose challenges, and the inherent noise in these systems can affect the performance and reliability of the models. These limitations underscore the need for ongoing advancements in both quantum hardware and simulators, enabling a reduction in computation times.

\section{Conclusions and Future Work}\label{sec:conclusions}

The integration of neuroscience and neuropsychology principles into quantum models offers a promising avenue for better emulating memory and learning processes. Investigating how these theories can be embedded into quantum frameworks could unlock significant advancements in quantum artificial intelligence, potentially leading to breakthroughs in the field.

Shifting focus to Online Learning (OL), where the learning algorithm receives a single sample at each time step, there is a critical need for novel algorithms capable of adapting quickly to changes while maintaining high performance. OL, particularly in scenarios involving concept drift, has become a topic of considerable interest in recent years, sparking ongoing debate within the research community due to its numerous unresolved challenges \cite{gama2014survey, krawczyk2017ensemble, losing2018incremental}. Although this manuscript does not explore OL in depth, future work could demonstrate the model's adaptability to real-time situations. This could involve leveraging Quantum Spiking Neural Networks for efficient real-time gradient calculations \cite{SNNTraining, SnnOL}, combined with the use of Quantum Long Short-Term Memory (QLSTM) networks to address issues such as catastrophic forgetting.

Finally, future research could explore the scalability of the proposed models to larger and more complex datasets, as well as their robustness in noisy or uncertain environments. Interestingly, one notable finding from this study is that the quantum models were able to achieve strong performance despite being trained on significantly smaller datasets. This data efficiency not only reduces training time—especially valuable when using quantum simulators with high computational demands—but also offers a strategic advantage in domains where labeled data is scarce. In such contexts, avoiding the need to generate synthetic data, a common yet sometimes controversial practice in data science, further underscores the applicability and practicality of quantum models.

Investigating the interplay between quantum computing and neuromorphic engineering could also lead to the development of innovative architectures that bridge the gap between hardware efficiency and cognitive functionality. Furthermore, interdisciplinary collaboration between neuroscience, artificial intelligence, and quantum computing holds the potential to uncover novel and compelling research directions, wherein these fields mutually inform and enhance each other, facilitating the creation of a unified mathematical model that integrates our understanding of brain function and applies this knowledge to the development of advanced algorithms.

\bibliographystyle{unsrt} 
\bibliography{references}  

\begin{thebibliography}{10}

\bibitem{WhenbraininspiredAImeetsAGI}
Lin Zhao, Lu~Zhang, Zihao Wu, Yuzhong Chen, Haixing Dai, Xiaowei Yu, Zhengliang Liu, Tuo Zhang, Xintao Hu, Xi~Jiang, Xiang Li, Dajiang Zhu, Dinggang Shen, and Tianming Liu.
\newblock When brain-inspired ai meets agi.
\newblock {\em Meta-Radiology}, 1(1):100005, 2023.

\bibitem{VQA-Classifiers}
Maria Schuld, Alex Bocharov, Krysta~M. Svore, and Nathan Wiebe.
\newblock Circuit-centric quantum classifiers.
\newblock {\em Phys. Rev. A}, 101:032308, 2020.

\bibitem{VQA-Neighbor}
Nathan Wiebe, Ashish Kapoor, and Krysta~M. Svore.
\newblock Quantum nearest-neighbor algorithms for machine learning.
\newblock {\em Quantum Information and Computation}, 15:318--358, 2015.

\bibitem{VQA-SVM}
D.~Anguita, Sandro Ridella, Fabio Rivieccio, and Rodolfo Zunino.
\newblock Quantum optimization for training support vector machines.
\newblock {\em Neural networks}, 16:763--770, 2003.

\bibitem{VQA_Unsupervised}
Esma A{\"i}meur, Gilles Brassard, and S{\'e}bastien Gambs.
\newblock Quantum speed-up for unsupervised learning.
\newblock {\em Machine Learning}, 90:261--287, 2013.

\bibitem{QRL-energies}
Eva Andrés, Manuel~Pegalajar Cuéllar, and Gabriel Navarro.
\newblock On the use of quantum reinforcement learning in energy-efficiency scenarios.
\newblock {\em Energies}, 15, 2022.

\bibitem{prefrontalcortex}
Earl~K. Miller and Jonathan~D. Cohen.
\newblock An integrative theory of prefrontal cortex function.
\newblock {\em Annual Review of Neuroscience}, 24(1):167--202, March 2001.

\bibitem{memory}
R.C. Atkinson and R.M. Shiffrin.
\newblock Human memory: A proposed system and its control processes.
\newblock In Kenneth~W. Spence and Janet~Taylor Spence, editors, {\em Human Memory: A Proposed System and its Control Processes.}, volume~2 of {\em Psychology of Learning and Motivation}, pages 89--195. Academic Press, 1968.

\bibitem{andersen2007hippocampus}
Per Andersen.
\newblock {\em The hippocampus book}.
\newblock Oxford university press, 2007.

\bibitem{olton1979hippocampus}
David~S Olton, James~T Becker, and Gail~E Handelmann.
\newblock Hippocampus, space, and memory.
\newblock {\em Behavioral and Brain sciences}, 2(3):313--322, 1979.

\bibitem{SNNTraining}
Jason~K. Eshraghian, Max Ward, Emre Neftci, Xinxin Wang, Gregor Lenz, Girish Dwivedi, Mohammed Bennamoun, Doo~Seok Jeong, and Wei~D. Lu.
\newblock Training spiking neural networks using lessons from deep learning, 2021.

\bibitem{cho-etal-2014-learning}
Kyunghyun Cho, Bart van Merri{\"e}nboer, Caglar Gulcehre, Dzmitry Bahdanau, Fethi Bougares, Holger Schwenk, and Yoshua Bengio.
\newblock Learning phrase representations using {RNN} encoder{--}decoder for statistical machine translation.
\newblock In Alessandro Moschitti, Bo~Pang, and Walter Daelemans, editors, {\em Proceedings of the 2014 Conference on Empirical Methods in Natural Language Processing ({EMNLP})}, pages 1724--1734, Doha, Qatar, October 2014. Association for Computational Linguistics.

\bibitem{hebb2005organization}
D.O. Hebb.
\newblock {\em The Organization of Behavior: A Neuropsychological Theory}.
\newblock Taylor \& Francis, 2005.

\bibitem{DLSNN}
Amirhossein Tavanaei, Masoud Ghodrati, Saeed~Reza Kheradpisheh, Timoth{\'{e}}e Masquelier, and Anthony Maida.
\newblock Deep learning in spiking neural networks.
\newblock {\em Neural Networks}, 111:47--63, 2019.

\bibitem{SnnOL}
Jesus~L. Lobo, Javier {Del Ser}, Albert Bifet, and Nikola Kasabov.
\newblock Spiking neural networks and online learning: An overview and perspectives.
\newblock {\em Neural Networks}, 121:88--100, 2020.

\bibitem{lapicque1907recherches}
LM~Lapicque.
\newblock Recherches quantitatives sur l’excitation electrique des nerfs.
\newblock {\em J. Physiol. Paris.}, 9:620--635, 1907.

\bibitem{SnnHdc}
Zhuowen Zou, Haleh Alimohamadi, Ali Zakeri, Farhad Imani, Yeseong Kim, M.~Hassan Najafi, and Mohsen Imani.
\newblock Memory-inspired spiking hyperdimensional network for robust online learning.
\newblock {\em Scientific Reports}, 12(1):7641, May 2022.

\bibitem{snnmuscle}
Kaushalya Kumarasinghe, Nikola Kasabov, and Denise Taylor.
\newblock Brain-inspired spiking neural networks for decoding and understanding muscle activity and kinematics from electroencephalography signals during hand movements.
\newblock {\em Scientific Reports}, 11(1):2486, Jan 2021.

\bibitem{navigationRL}
Andrea Banino, Caswell Barry, Benigno Uria, Charles Blundell, Timothy Lillicrap, Piotr Mirowski, Alexander Pritzel, Martin~J. Chadwick, Thomas Degris, Joseph Modayil, Greg Wayne, Hubert Soyer, Fabio Viola, Brian Zhang, Ross Goroshin, Neil Rabinowitz, Razvan Pascanu, Charlie Beattie, Stig Petersen, Amir Sadik, Stephen Gaffney, Helen King, Koray Kavukcuoglu, Demis Hassabis, Raia Hadsell, and Dharshan Kumaran.
\newblock Vector-based navigation using grid-like representations in artificial agents.
\newblock {\em Nature}, 557(7705):429--433, May 2018.

\bibitem{lstmlongtermdependencies}
Y.~Bengio, P.~Simard, and P.~Frasconi.
\newblock Learning long-term dependencies with gradient descent is difficult.
\newblock {\em IEEE Transactions on Neural Networks}, 5:157--166, 1994.

\bibitem{lstmhandwritting}
Alex Graves, Marcus Liwicki, Santiago Fernández, Roman Bertolami, Horst Bunke, and Jürgen Schmidhuber.
\newblock A novel connectionist system for unconstrained handwriting recognition.
\newblock {\em IEEE Transactions on Pattern Analysis and Machine Intelligence}, 31:855--868, 2009.

\bibitem{lstmclassification}
Alex Graves and J{\"u}rgen Schmidhuber.
\newblock Framewise phoneme classification with bidirectional lstm and other neural network architectures.
\newblock {\em Neural networks : the official journal of the International Neural Network Society}, 18:602--610, 2005.

\bibitem{Lstmsolve}
Sepp Hochreiter and Jürgen Schmidhuber.
\newblock Lstm can solve hard long time lag problems.
\newblock {\em Advances in Neural Information Processing Systems}, 9:473--479, 1996.

\bibitem{lstm}
Anthony Triche, Anthony~S. Maida, and Ashok Kumar.
\newblock Exploration in neo-hebbian reinforcement learning: Computational approaches to the exploration–exploitation balance with bio-inspired neural networks.
\newblock {\em Neural Networks}, 151:16--33, 2022.

\bibitem{VQA_for_QNN}
Antonio Macaluso, Luca Clissa, Stefano Lodi, and Claudio Sartori.
\newblock A variational algorithm for quantum neural networks.
\newblock In {\em Computational Science -- ICCS 2020}, pages 591--604. Springer International Publishing, 2020.

\bibitem{QML}
Marcello Benedetti, Erika Lloyd, Stefan Sack, and Mattia Fiorentini.
\newblock Parameterized quantum circuits as machine learning models.
\newblock {\em Quantum Science and Technology}, 4(4):043001, nov 2019.

\bibitem{QNN-layers}
Chen Zhao and Xiao-Shan Gao.
\newblock Qdnn: deep neural networks with quantum layers.
\newblock {\em Quantum Machine Intelligence}, 3:15, 2021.

\bibitem{QuantumGenerativeLearning}
Jinkai Tian, Xiaoyu Sun, Yuxuan Du, Shanshan Zhao, Qing Liu, Kaining Zhang, Wei Yi, Wanrong Huang, Chaoyue Wang, Xingyao Wu, Min-Hsiu Hsieh, Tongliang Liu, Wenjing Yang, and Dacheng Tao.
\newblock Recent advances for quantum neural networks in generative learning.
\newblock {\em IEEE Transactions on Pattern Analysis and Machine Intelligence}, 45(10):12321--12340, 2023.

\bibitem{QuantumSelfAttention}
Guangxi Li, Xuanqiang Zhao, and Xin Wang.
\newblock Quantum self-attention neural networks for text classification.
\newblock {\em Science China Information Sciences}, 67(4):142501, 2024.

\bibitem{QMLBook}
Peter Wittek.
\newblock {\em Quantum Machine Learning: What Quantum Computing means to data mining}.
\newblock Elsevier, 2014.

\bibitem{Expanding-DataEncoding}
Manuela Weigold, Johanna Barzen, Frank Leymann, and Marie Salm.
\newblock Expanding data encoding patterns for quantum algorithms.
\newblock In {\em 2021 IEEE 18th International Conference on Software Architecture Companion (ICSA-C)}, pages 95--101, 2021.

\bibitem{Catastrophic}
Michael McCloskey and Neal~J. Cohen.
\newblock Catastrophic interference in connectionist networks: The sequential learning problem.
\newblock In Gordon~H. Bower, editor, {\em Catastrophic Interference in Connectionist Networks: The Sequential Learning Problem}, volume~24 of {\em Psychology of Learning and Motivation}, pages 109--165. Academic Press, 1989.

\bibitem{continualLearning}
Friedemann Zenke, Ben Poole, and Surya Ganguli.
\newblock Continual learning through synaptic intelligence.
\newblock In Doina Precup and Yee~Whye Teh, editors, {\em Proceedings of the 34th International Conference on Machine Learning}, volume~70 of {\em Proceedings of Machine Learning Research}, pages 3987--3995. PMLR, 06--11 Aug 2017.

\bibitem{ProgressiveNeuralNetworks}
Andrei~A. Rusu, Neil~C. Rabinowitz, Guillaume Desjardins, Hubert Soyer, James Kirkpatrick, Koray Kavukcuoglu, Razvan Pascanu, and Raia Hadsell.
\newblock Progressive neural networks.
\newblock {\em CoRR}, abs/1606.04671, 2016.

\bibitem{GenerativeReplay}
Hanul Shin, Jung~Kwon Lee, Jaehong Kim, and Jiwon Kim.
\newblock Continual learning with deep generative replay.
\newblock {\em CoRR}, abs/1705.08690, 2017.

\bibitem{overcomingcatastrhophic}
James Kirkpatrick, Razvan Pascanu, Neil~C. Rabinowitz, Joel Veness, Guillaume Desjardins, Andrei~A. Rusu, Kieran Milan, John Quan, Tiago Ramalho, Agnieszka Grabska{-}Barwinska, Demis Hassabis, Claudia Clopath, Dharshan Kumaran, and Raia Hadsell.
\newblock Overcoming catastrophic forgetting in neural networks.
\newblock {\em CoRR}, abs/1612.00796, 2016.

\bibitem{snntorch}
Jason~K Eshraghian, Max Ward, Emre Neftci, Xinxin Wang, Gregor Lenz, Girish Dwivedi, Mohammed Bennamoun, Doo~Seok Jeong, and Wei~D Lu.
\newblock Training spiking neural networks using lessons from deep learning.
\newblock {\em Proceedings of the IEEE}, 111(9):1016--1054, 2023.

\bibitem{gama2014survey}
Jo{\~a}o Gama, Indr{\.e} {\v{Z}}liobait{\.e}, Albert Bifet, Mykola Pechenizkiy, and Abdelhamid Bouchachia.
\newblock A survey on concept drift adaptation.
\newblock {\em ACM computing surveys (CSUR)}, 46(4):1--37, 2014.

\bibitem{krawczyk2017ensemble}
Bartosz Krawczyk, Leandro~L Minku, Jo{\~a}o Gama, Jerzy Stefanowski, and Micha{\l} Wo{\'z}niak.
\newblock Ensemble learning for data stream analysis: A survey.
\newblock {\em Information Fusion}, 37:132--156, 2017.

\bibitem{losing2018incremental}
Viktor Losing, Barbara Hammer, and Heiko Wersing.
\newblock Incremental on-line learning: A review and comparison of state of the art algorithms.
\newblock {\em Neurocomputing}, 275:1261--1274, 2018.

\end{thebibliography}

\end{document}